\documentclass[journal]{IEEEtran}
\IEEEoverridecommandlockouts
\usepackage{cite}
\usepackage{enumitem}
\usepackage{amsmath,amssymb,amsfonts,amsthm}
\usepackage{algorithm}
\usepackage{algpseudocode}
\usepackage{graphicx}
\usepackage{textcomp}
\usepackage{xcolor}
\usepackage{relsize}
\usepackage{comment}
\usepackage{bigints}
\usepackage{tabularx}
\usepackage{url} 
\usepackage{scalerel}
\usepackage{tikz}
\usepackage{acronym}
\usepackage{setspace}
\usepackage{multirow}
\usepackage[hidelinks]{hyperref} 
\ifCLASSOPTIONcompsoc
    \usepackage[caption=false, font=normalsize, labelfont=sf, textfont=sf]{subfig}
\else
\usepackage[caption=false, font=normalsize]{subfig}
\fi

\def\BibTeX{{\rm B\kern-.05em{\sc i\kern-.025em b}\kern-.08em
    T\kern-.1667em\lower.7ex\hbox{E}\kern-.125emX}}
\usetikzlibrary{svg.path}
\definecolor{orcidlogocol}{HTML}{A6CE39}
\tikzset{
    orcidlogo/.pic={
        \fill[orcidlogocol] svg{M256,128c0,70.7-57.3,128-128,128C57.3,256,0,198.7,0,128C0,57.3,57.3,0,128,0C198.7,0,256,57.3,256,128z};
        \fill[white] svg{M86.3,186.2H70.9V79.1h15.4v48.4V186.2z}
        svg{M108.9,79.1h41.6c39.6,0,57,28.3,57,53.6c0,27.5-21.5,53.6-56.8,53.6h-41.8V79.1z M124.3,172.4h24.5c34.9,0,42.9-26.5,42.9-39.7c0-21.5-13.7-39.7-43.7-39.7h-23.7V172.4z}
        svg{M88.7,56.8c0,5.5-4.5,10.1-10.1,10.1c-5.6,0-10.1-4.6-10.1-10.1c0-5.6,4.5-10.1,10.1-10.1C84.2,46.7,88.7,51.3,88.7,56.8z};
    }
}

\newcommand\orcidicon[1]{\href{https://orcid.org/#1}{\mbox{\scalerel*{
                \begin{tikzpicture}[yscale=-1,transform shape]
                \pic{orcidlogo};
                \end{tikzpicture}
            }{|}}}}
\makeatletter
\def\ignorecitefornumbering#1{%
     \begingroup
         \@fileswfalse
         #1
    \endgroup
}
\makeatother
\newacro{IoT}[IoT]{internet-of-things}
\newacro{LEO}[LEO]{low Earth orbit}
\newacro{GEO}[GEO]{geostationary orbit}
\newacro{LoS}[LoS]{line-of-sight}
\newacro{GS}[GS]{ground station}
\newacro{AF}[AF]{amplify-and-forward}
\newacro{MRC}[MRC]{maximal ratio combining}
\newacro{SNR}[SNR]{signal-to-noise ratio}
\newacro{SINR}[SINR]{signal-to-interference-plus-noise ratio}
\newacro{LPWA}[LPWA]{low-power wide-area}
\newacro{LoRaWAN}[LoRaWAN]{long range wide area network}
\newacro{NB-IoT}[NB-IoT]{narrowband-IoT}
\newacro{HSTN}[HSTN]{hybrid satellite-terrestrial network}
\newacro{NOMA}[NOMA]{non-orthogonal multiple access}
\newacro{SS}[SS]{single satellite}
\newacro{AWGN}[AWGN]{additive white Gaussian noise}
\newacro{PDF}[PDF]{probability density function}
\newacro{CDF}[CDF]{cumulative distribution function}
\newacro{MGF}[MGF]{moment generating function}
\newacro{SR}[SR]{shadowed-Rician}
\newacro{EIRP}[EIRP]{equivalent isotropically radiated power}
\newacro{G/T}[G/T]{antenna gain-to-noise-temperature}
\newacro{OP}[OP]{outage probability}
\newacro{DF}[DF]{decode-and-forward}
\newacro{LPWAN}[LPWAN]{low power wide area network}
\newacro{SIC}[SIC]{successive interference cancellation}
\newacro{BPP}[BPP]{binomial point process}
\newacro{CM}[CM]{capture model}
\newacro{MAC}[MAC]{medium access control}
\newacro{AWGN}[AWGN]{additive white Gaussian noise}
\newacro{DtS-IoT}[DtS-IoT]{direct-to-satellite IoT}
\newacro{CSI}[CSI]{channel state information }
\newacro{PHY}[PHY]{physical}
\newacro{GA}[GA]{Genetic algorithm}

\newcommand{\p}{\mathbb{P}}
\newcommand{\gt}{\gamma_{\text{th}}}
\newcommand\txtblue[1]{{\color{black}#1}}

\theoremstyle{plain}

\begin{document}
\title{Performance Analysis of LEO Satellite-Based IoT Networks in the Presence of Interference\\
}

\author{Ayush Kumar Dwivedi${\textsuperscript{\orcidicon{0000-0003-2395-6526}}}$, \IEEEmembership{Student Member, IEEE}, Sachin Chaudhari${\textsuperscript{\orcidicon{0000-0003-1923-0925}}}$, \IEEEmembership{Senior Member, IEEE}, \\ Neeraj Varshney${\textsuperscript{\orcidicon{0000-0003-0752-2103}}}$, \IEEEmembership{Senior Member, IEEE}, Pramod K. Varshney${\textsuperscript{\orcidicon{0000-0003-4504-5088}}}$, \IEEEmembership{Life Fellow, IEEE}

\thanks{Ayush Kumar Dwivedi (\textit{Corresponding author}) and Sachin Chaudhari are with the Signal Processing and Communication Research Center at International Institute of Information Technology, Hyderabad 500032, India (e-mail: ayush.dwivedi@research.iiit.ac.in; sachin.c@iiit.ac.in).}
\thanks{Neeraj Varshney is with the Wireless Networks Division, National Institute of Standards and Technology, Gaithersburg, MD 20899 USA (e-mail: neerajv@ieee.org).}
\thanks{Pramod K. Varshney is with the Department of Electrical Engineering and Computer Science, Syracuse University, Syracuse, NY 13244 USA (e-mail: varshney@syr.edu).}
}

\maketitle

\begin{abstract}
This paper presents a star-of-star topology for internet-of-things (IoT) networks using mega low-Earth-orbit constellations. The proposed topology enables IoT users to broadcast their sensed data to multiple satellites simultaneously over a shared channel, which is then relayed to the ground station (GS) using amplify-and-forward relaying. The GS coherently combines the signals from multiple satellites using maximal ratio combining. To analyze the performance of the proposed topology in the presence of interference, a comprehensive outage probability (OP) analysis is performed, assuming imperfect channel state information at the GS. The paper employs stochastic geometry to model the random locations of satellites, making the analysis general and independent of any specific constellation. Furthermore, the paper examines successive interference cancellation (SIC) and capture model (CM)-based decoding schemes at the GS to mitigate interference. The average OP for the CM-based scheme and the OP of the best user for the SIC scheme are derived analytically. The paper also presents simplified expressions for the OP under a high signal-to-noise ratio (SNR) assumption, which are utilized to optimize the system parameters for achieving a target OP. The simulation results are consistent with the analytical expressions and provide insights into the impact of various system parameters, such as mask angle, altitude, number of satellites, and decoding order. The findings of this study demonstrate that the proposed topology can effectively leverage the benefits of multiple satellites to achieve the desired OP and enable burst transmissions without coordination among IoT users, making it an attractive choice for satellite-based IoT networks.
\end{abstract}


\section{Introduction}
\label{intro}
The emerging \ac{IoT} networks aim to connect a large number of devices and sensors over a wide range of applications such as smart cities, e-healthcare, marine \ac{IoT}, and connected vehicles. Despite significant improvements in terrestrial wireless systems, providing coverage at remote locations remains a challenge, with only 25\% of the world's landmass having terrestrial connectivity \cite{coverage}. The intrinsic broadcasting capability of satellite systems makes them a viable solution for delivering truly ubiquitous service to \ac{IoT} networks often deployed remotely over large areas \cite{survey}. A recent 3GPP Rel-17 study-item has also specified the support required for satellite-based NB-IoT/eMTC networks \cite{standard_1}. Recently, many \ac{LEO} satellite constellations, e.g., Starlink-SpaceX, OneWeb, Kuiper, Telesat, etc., have been launched. They can potentially serve the remotely deployed \ac{IoT} networks with multiple \ac{LEO} satellites in the visible range for a large fraction of the Earth's surface \cite{consti}. Moreover, they have relatively low propagation delay at lower powers when compared to the legacy \ac{GEO} satellites.\looseness=-1

The previous studies in \cite{survey1,survey2,ingr} have identified several architectural challenges and enabling solutions for satellite-based \ac{IoT} networks. Specifically, they suggest that upgrades are needed at the \ac{PHY} and \ac{MAC} layers to include non-orthogonal and non-pure ALOHA-based approaches, as well as the exploration of computationally simple random access schemes and novel topologies for massive \ac{IoT} connectivity. In this paper, we address some of these requirements by proposing a star-of-star topology-based satellite-\ac{IoT} network. In the past, star topologies have been used in terrestrial \ac{LPWAN} technologies like Sigfox and \ac{LoRaWAN} \cite{lora2,lora1}, but their performance has not been explored for satellite-based \ac{IoT} networks as done in this paper. Unlike previous studies \cite{stochastic1,stochastic2,geo1}, where either uplink or downlink performance is analyzed for simple channel models, we investigate the end-to-end performance using more realistic channel fading models for satellite communication. We also consider the impact of random access with multiple \ac{IoT} users transmitting simultaneously as well as the fact that longer distances between \ac{IoT} users and satellites lead to higher path loss. To address the above factors, we propose a more general system that leverages multiple satellites for improved performance.\looseness=-1

Our proposed topology also addresses the energy consumption challenges associated with \ac{IoT} users, which have limited energy resources. Unlike cellular networks, where significant energy is consumed in listening while in the idle state and in synchronization with the base station for continuous coverage, \ac{IoT} devices only require intermittent coverage for data transmission. The proposed topology facilitates burst mode of operation that allows \ac{IoT} devices to transmit data at specific intervals and sleep thereafter, without requiring synchronization and routing. This ensures minimal computational complexity for \ac{IoT} users and all the processing is shifted to the \ac{GS}. The presented results focus on the \ac{PHY} layer aspects with a particular emphasis on topology and performance analysis.\looseness=-1
\begin{table*}[t!]
\centering
\caption{Comparison of work done in this paper with other papers in the literature}
\label{Lit}
\begin{tabular}{l|ccccccccc}
\hline \hline
Scope/Reference & \ignorecitefornumbering{\cite{lora2,lora1}} & \ignorecitefornumbering{\cite{stochastic1,stochastic2}} & \ignorecitefornumbering{\cite{geo1}} & \ignorecitefornumbering{\cite{df1}} & \ignorecitefornumbering{\cite{af_0}} & \ignorecitefornumbering{\cite{af_sic_2}} & \ignorecitefornumbering{\cite{niloofarDownlink,Niloofar2}} &  \ignorecitefornumbering{\cite{stochastic3}} & This paper\\
\hline
Satellite channel model  &  & \checkmark & \checkmark &  & \checkmark &  &  & \checkmark & \checkmark \\

Direct-access AF topology & & & & & \checkmark & \checkmark &  &  & \checkmark \\

Multiple satellites/relays & & & & \checkmark & & \checkmark &  &  & \checkmark \\

Interference from other users & \checkmark & \checkmark &  & \checkmark &  & \checkmark & \checkmark &  & \checkmark \\

Combining at the GS  &  &  &  & \checkmark &  & \checkmark &  &  & \checkmark \\

Random location of satellites &  & \checkmark & 
 &  &  &  & \checkmark & \checkmark & \checkmark \\

SIC decoding at GS & \checkmark &  &  &  &  &  &  &  & \checkmark \\

OP analysis on end-to-end SNR/SINR & \checkmark & \checkmark & \checkmark & \checkmark &  & \checkmark & \checkmark & \checkmark & \checkmark \\

Imperfect channel estimates  &  &  &  &  &  &  &  &  & \checkmark \\

Asymptotic analysis on SNR  &  &  & \checkmark & \checkmark & \checkmark & \checkmark &  &  & \checkmark \\

Analysis on effect of system parameters &  & \checkmark & \checkmark & \checkmark &  &  & \checkmark & \checkmark & \checkmark \\
\hline\hline
\end{tabular}
\vspace{0.2cm}
\end{table*}
\subsection{Related Work}
In a star-of-star topology, the access between the node and the relay can be direct or indirect, where the satellites can act as a relay/repeater between the node and the server. However, \ac{DtS-IoT} has recently gained traction because of its ease of deployment \cite{direct}. In \cite{loraDirectAccess}, it is shown that \ac{LPWAN} technologies can be configured for realising \ac{DtS-IoT} communication. Moreover, some manufacturers' low-cost, battery-powered development kits have also provided the impetus to DtS-IoT using LEO satellites \cite{lacuna,astrocast,orbcomm}. The feasibility of \ac{DtS-IoT} has been established by the link budget analysis carried out on \ac{IoT} users of various power classes by different companies in a recent 3GPP study-item \cite{standard_1}.

In \ac{DtS-IoT}, the satellites can act as regenerative (\ac{DF}) or transparent (\ac{AF}) relays. In \cite{df1}, the performance of a \ac{DF} relaying network with randomly distributed interferers is analyzed under Rayleigh and Nakagami-\textit{m} fading channels. In \cite{nikhil}, the performance of selective-\ac{DF} relaying for \ac{DtS-IoT} network has been analysed for \ac{LEO} satellites. But, the complexity of \ac{AF} relaying with a fixed gain is less than \ac{DF} relaying \cite{ray}, and has been preferred in \cite{standard_1}. The performance of topologies employing \ac{AF} relaying has been widely studied for various systems. For example, in \cite{af_0}, the performance of a hybrid satellite-terrestrial cooperative network consisting of a single \ac{AF} relay has been analyzed for generalized fading. In \cite{af_1}, \cite{af_2}, \cite{af_3}, and \cite{af_4}, the performance of an \ac{AF} system with multiple relays and \ac{MRC} at the destination has been analysed for Rayleigh, Nakagami-\textit{m}, Rician, and shadowed-Rician faded channels, respectively. Additionally, co-channel interference has been included in the performance analysis of relay-based topologies in \cite{afInt_1}, and \cite{afInt_2} for Rayleigh and mixed-Rayleigh-Rician fading channels, respectively. However, this interference can be mitigated using various cancellation techniques. In this context, several interference mitigation techniques for both relay and non-relay systems have also been discussed in the literature. A \ac{NOMA} inspired system with a single \ac{AF} relay and a decoding scheme using the signal from two consecutive time slots is analysed in \cite{af_sic_1}. Similarly a multi-source, multi-relay system with opportunistic interference cancellation using adaptive \ac{AF}/\ac{DF} is analysed in \cite{af_sic_2}. In both \cite{af_sic_1} and \cite{af_sic_2}, performance is analysed under the Rayleigh fading assumption. In \cite{lora2,lora1}, non-relay terrestrial \ac{IoT} communication systems have been studied. A \ac{LoRaWAN} like system with and without fading is considered in \cite{lora2} and \cite{lora1}, respectively, to show that the \ac{CM} and \ac{SIC}-based decoding schemes can perform better than the traditional ALOHA schemes. In \ac{CM}, the strongest received signal can be decoded successfully despite interference if the \ac{SINR} is greater than the threshold. Whereas in \ac{SIC}, decoding is performed in the order of \ac{SINR}s while cancelling the interference at each step. While \ac{NOMA}-based \ac{IoT} network may seem attractive for satellite-based systems as it has more degrees of freedom \cite{satnoma1,satnoma2}, it is not practical as it will require sophisticated power control algorithms and continuous communication with the ground station, which is not feasible for \ac{IoT} networks with a large number of devices and limited computational and battery resources.
\begin{figure*}[t!]
\centering
\setkeys{Gin}{width=\linewidth}
\begin{tabularx}{\linewidth}{XXXX}
    \centering
    \includegraphics[width=2in,trim={3.9cm 2.1cm 3.2cm 1.5cm},clip]{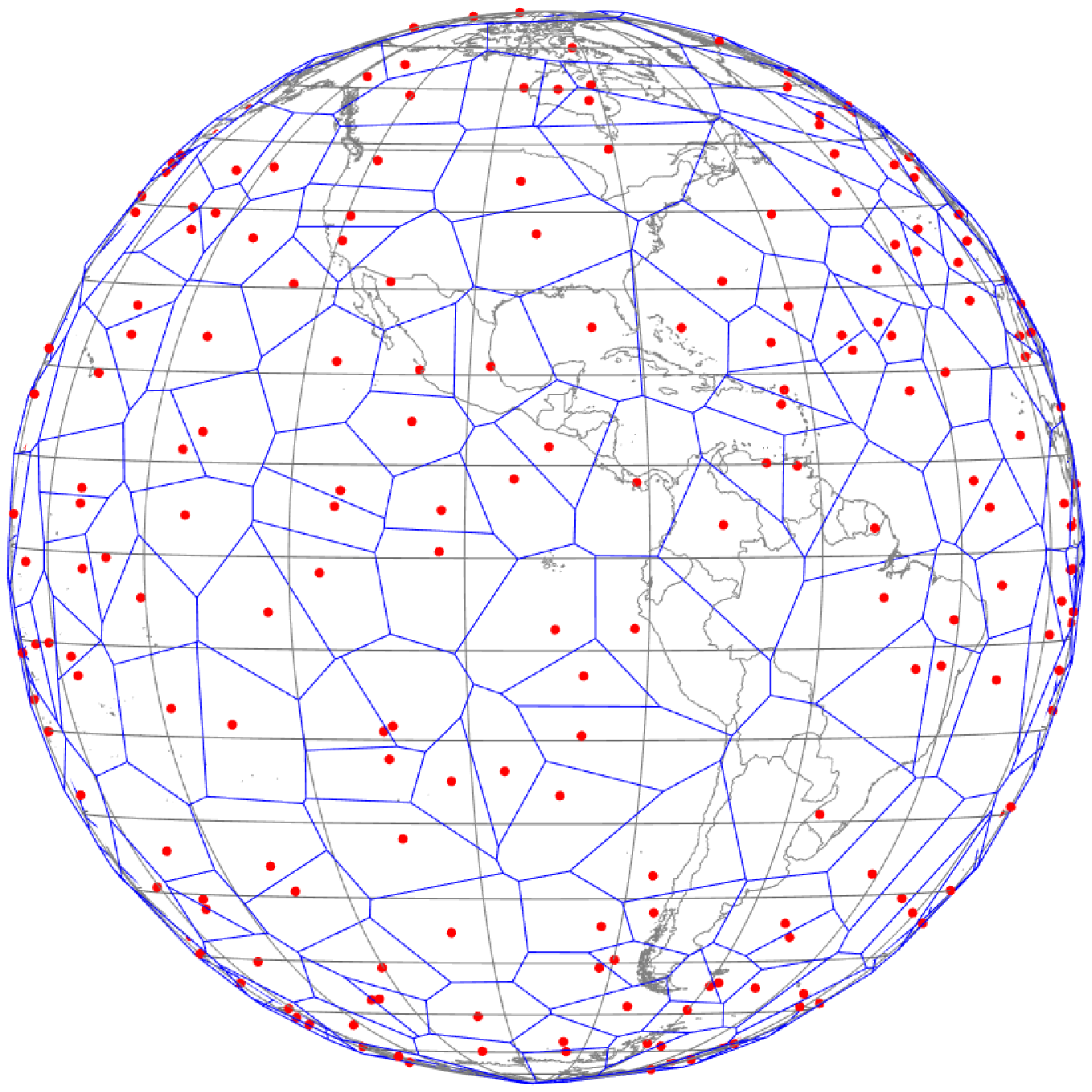}
    \caption{An example constellation of size $K=720$ satellites where the satellites are distributed on a spherical surface following a \ac{BPP}. The Voronoi diagrams represent the area to which the enclosed satellite is the nearest.}
    \label{constellation}
&
    \centering
    \includegraphics[width=0.99\linewidth,trim={0.4cm 0.5cm 0.4cm 0.3cm},clip]{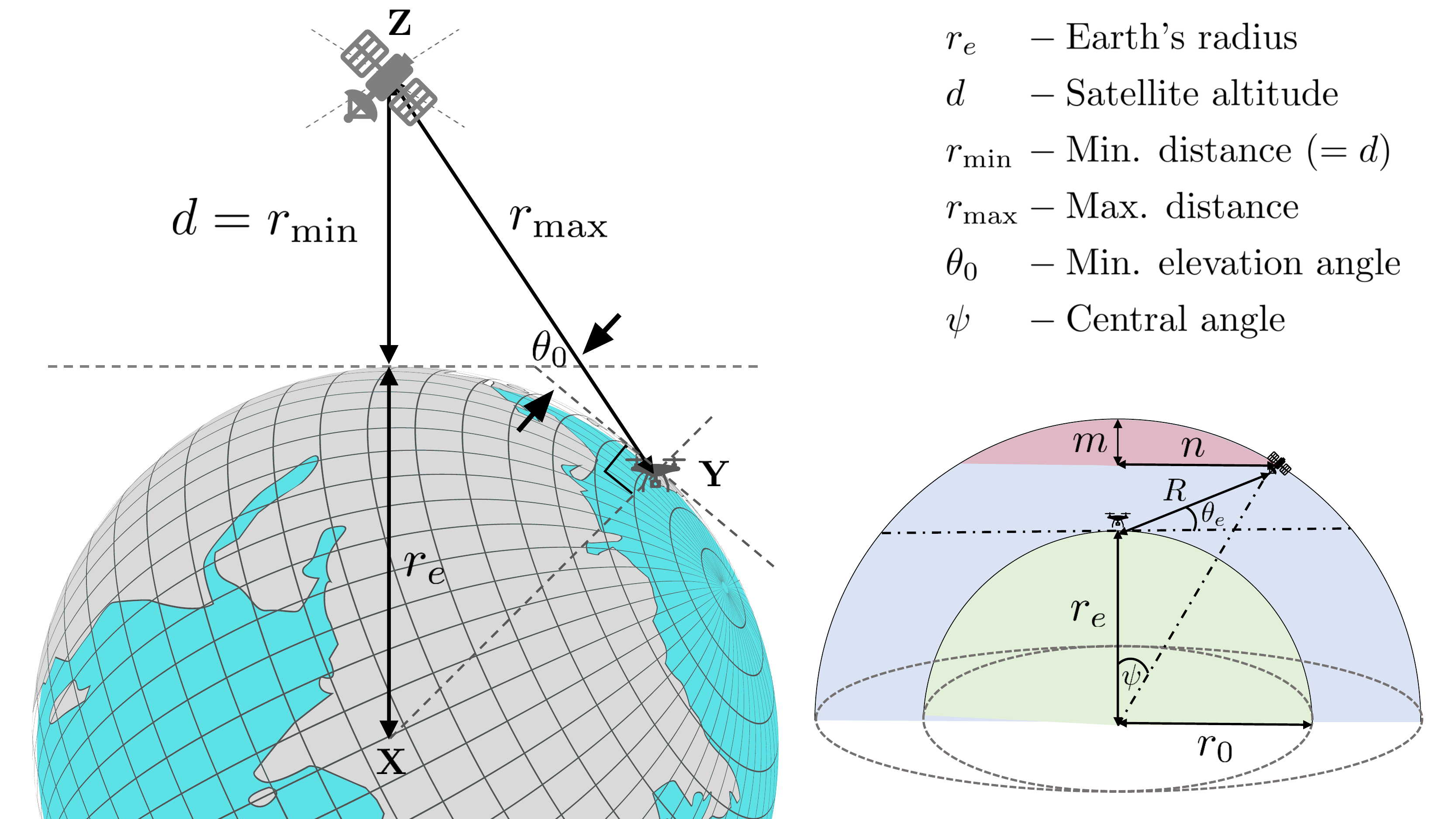}
    \caption{A geometric representation of the elevation angle $\theta_e$ and slant range $R$ between a particular satellite and the IoT user being decoded at \ac{GS} (maximum slant range $r_{\text{max}}$ is obtained when $\theta_e = \theta_0$).}
    \label{sm}
\end{tabularx}
\end{figure*}

Recently many mega-\ac{LEO} constellations have been launched, and more are being proposed for deployment in the near future. For these, performance analysis done using satellite locations based on their orbit simulations can not be generalized for any new constellation in future. Hence, to generalize the analysis for any constellation, tools from stochastic geometry have been used in recent literature where satellites are assumed to be randomly located around the Earth \cite{niloofarDownlink,Niloofar2,stochastic1,stochastic2, stochastic3}.
In \cite{stochastic1}, coverage and throughput performance for the uplink of a satellite-based-\ac{IoT} network has been presented using an empirical channel model representing path-loss and large-scale fading. In \cite{stochastic2}, a fine-grained analysis has been given for the downlink of a \ac{LEO} satellites-based mmWave relay network. The satellites are assumed to be uniformly distributed on a spherical cap around the Earth, and meta-distribution of the \ac{SINR} is used for performance evaluation. In \cite{niloofarDownlink,Niloofar2}, a theoretical analysis of downlink coverage and rate in a \ac{LEO} constellation is presented. The satellites are modelled as a \ac{BPP} on a sphere, and the users are located on the Earth's surface. Expressions for statistical characteristics of range and number of visible satellites have been derived along with the notion of an effective number of satellites to suppress the performance mismatch between the practical and random constellations. In \cite{stochastic3}, the performance of an \ac{LEO} satellite network with ground-based gateways acting as relays has been compared with a fiber-connected network to demonstrate the coverage gain for rural and remote areas using \ac{LEO} satellites. In the context of \ac{GEO} satellites, \cite{geo1} analyzes the performance of downlink channels for randomly located users in single and multibeam areas.

\subsection{Contributions of the paper}
Keeping in mind the requirements of \ac{IoT}, e.g., simple random access, novel topology, and the results from existing literature, this paper explores a star-of-star topology for satellite-based \ac{IoT} networks, which can leverage the benefits of multiple satellites in the visible range. The \ac{IoT} network is envisioned as the subscriber of one of the many services offered by mega-\ac{LEO} constellations. The satellites are assumed to be randomly distributed as a \ac{BPP} on a sphere around the Earth such that a user can see a satellite only if its elevation angle is greater than the mask angle. Multiple satellites simultaneously listen to the broadcast information from multiple \ac{IoT} users and forward it to the \ac{GS} using fixed-gain \ac{AF}. It is assumed that \ac{IoT} users wake up to offload the sensed data to all the visible satellites without any prior coordination and sleep again. This way, the processing is kept simple and energy-efficient for the \ac{IoT} users, and all the complexity is moved to the satellites and the \ac{GS}. Since many \ac{IoT} users are assumed to transmit simultaneously, this work considers \ac{CM} and SIC-based decoding at the \ac{GS}.

In \cite{af_1,af_2,af_3,af_4}, performance analysis was done without considering any interference at the relays as opposed to this paper. Similarly, in \cite{afInt_1,afInt_2}, although co-channel interference was considered, only a single relay was employed, different from the multiple satellite relay architecture considered in this paper. Neither \cite{lora1} nor \cite{lora2} considered a relay system with fading in the propagation environment while analyzing decoding schemes as done in this paper. Moreover, all the above papers neither specifically consider satellites as relays in their performance analysis, nor do they include any information about the location of the sources and relays.

In \cite{niloofarDownlink,Niloofar2,stochastic1,stochastic2,geo1,stochastic3}, the coverage and rate analysis is limited to only single link (either uplink or downlink) performance using a single-serving satellite. Also, no mask elevation angle has been considered to define the visibility of a satellite. In our previous preliminary work \cite{pimrc}, a similar topology as this paper was employed, but the analysis was limited to scenarios with single-user and no interference. Compared to \cite{pimrc}, the analysis in this work is extended to scenarios with multiple interfering users and channels with imperfect knowledge. Also, in \cite{pimrc}, all the visible satellites were considered at fixed locations. However, in this work, the performance of the employed topology is analyzed with different decoding schemes using stochastic geometry to generalize the analysis for any \ac{LEO} satellite constellation. Further, the system parameters like the number of devices, satellites, and mask angle are optimized in this paper to obtain an optimal region of operation. Table \ref{Lit} provides a comprehensive comparison of the aspects covered in the key previous works with respect to the aspects covered in this paper.
\begin{table*}[t!]
\centering \normalsize
\caption{Summary of Symbols and Notations}
\begin{tabular}{ c | c | c | c }
\hline\hline \label{notations}
\textbf{Symbol} & \textbf{Description} & \textbf{Symbol} & \textbf{Description} \\ 
\hline
$U$ & Number of IoT users & $S$ & Number of satellites used for AF \\
$K$ & Number of satellites in constellation & $K_{\text{vis}}$ & Number of visible satellites\\
$r_e$ & Radius of Earth (6371 km) & $d$ & Altitude of satellite \\  
$R_{us}$ & Distance between the $u^\text{th}$ user and $s^\text{th}$ satellite & $R_g$ & Distance between any satellite and GS \\
$\theta_e$ & Elevation angle & $\theta_0$ & Minimum elevation angle or mask angle \\
$P_u$ & Transmit power of the $u^\text{th}$ user & $P_s$ & Transmit power of the $s^\text{th}$ satellite \\
$\sigma_n^2$ & AWGN power at satellite for $u^\text{th}$ user & $\sigma^2_w$ & AWGN power at \ac{GS} for $s^\text{th}$ satellite \\
$\mathcal{R}$ & Target rate & $B$ & Available bandwidth \\
$h_{us}$ & Channel coefficient of $(u-s)$ user-satellite pair &  $g_s$ & Channel coefficient of $s^\text{th}$ satellite\\
$m,b,\Omega$ & Parameters of Shadowed Rician channel & $\gt$ & SINR threshold \\
$\alpha$ & Path loss exponent & $P_{\text{out}}$ & Outage probability\\
\hline\hline
\end{tabular}
\end{table*}

The specific contributions of this paper are as follows:
\begin{enumerate}
    \item A star-of-star topology is employed for satellite-based \ac{IoT} networks, which can leverage the benefits of multiple satellites in the visible range.
    \item The statistical characteristics of the range and the number of visible satellites for a given mask angle are derived in closed form, which is lacking in the existing literature. These characteristics arise from stochastic modelling and are crucial to finding the \ac{OP}.
    %
    \item The exact expression for the average \ac{OP} of a user at the \ac{GS} is derived for \ac{CM} and the \ac{OP} for the best user is derived for \ac{SIC} under the assumption that perfect \ac{CSI} is not known at the \ac{GS}. The derived theoretical results are validated with Monte-Carlo simulations.
    \item  Asymptotic expressions of \ac{OP} under high \ac{SNR} assumption are obtained, which are much simpler to comprehend and do not include any integrals. \txtblue{The proposed topology is demonstrated to achieve a diversity order equal to the number of satellites used for \ac{AF} in case of no-interference and perfect \ac{CSI}. However, the \ac{OP} attains a floor when there is interference and errors due to imperfect \ac{CSI}.} The asymptotic expressions are further utilized to optimize the system parameters like the number of devices, satellites, and mask angle to obtain the optimal region of operation achieving a target \ac{OP}.
    \item The effect of various key design parameters like the number of satellites, altitude, elevation angle and user ordering in \ac{SIC} on the OP are analyzed.
\end{enumerate}

The rest of the paper is organised as follows: Section \ref{sys} presents the detailed system model, and Section \ref{R_Kvis} discusses the statistical characteristics of the range and number of visible satellites. The exact \ac{OP} derivations for both \ac{CM} and \ac{SIC} decoding schemes are derived in Section \ref{OPDerivation} and the asymptotic expressions are derived in Section \ref{Sec:AsympOP}. The results and the associated discussions are presented in Section \ref{Results}, followed by the conclusion in Section \ref{sec:conc}.
%
\vspace{0.2cm}
\section{System Description}
\label{sys}
As shown in Fig. \ref{constellation}, a total of $K$ satellites are assumed to be distributed uniformly around the Earth at an altitude $d$ km such that they form a \ac{BPP} on a sphere of radius $r_e+d$, where $r_e$ is the radius of the Earth. The users are assumed to be located on the Earth's surface. A satellite is considered visible and can receive a signal from a user only if its elevation angle $\theta_e$ w.r.t user's location is greater than a minimum elevation or mask angle $\theta_0$—any satellite for which $\theta_e < \theta_0$ is considered invisible to the user. As shown in Fig. \ref{sm}, the distance between a user and a satellite will be minimum when the satellite is at maximum elevation $\theta_e = 90^\circ$ w.r.t the user. This minimum distance $r_{\text{min}}$ equals the altitude $d$ at which all the satellites in the constellation are deployed. Similarly, the distance between a user and a satellite is maximum when the satellite is at an elevation angle $\theta_e = \theta_0$ w.r.t to the user. The maximum distance for a fixed $\theta_0$ can be derived as shown in Appendix \ref{proof:range}. All the notations followed in this paper are defined in Table \ref{notations}.\looseness=-1

\begin{figure}[t!]
\centering
\includegraphics[width=3.1in]{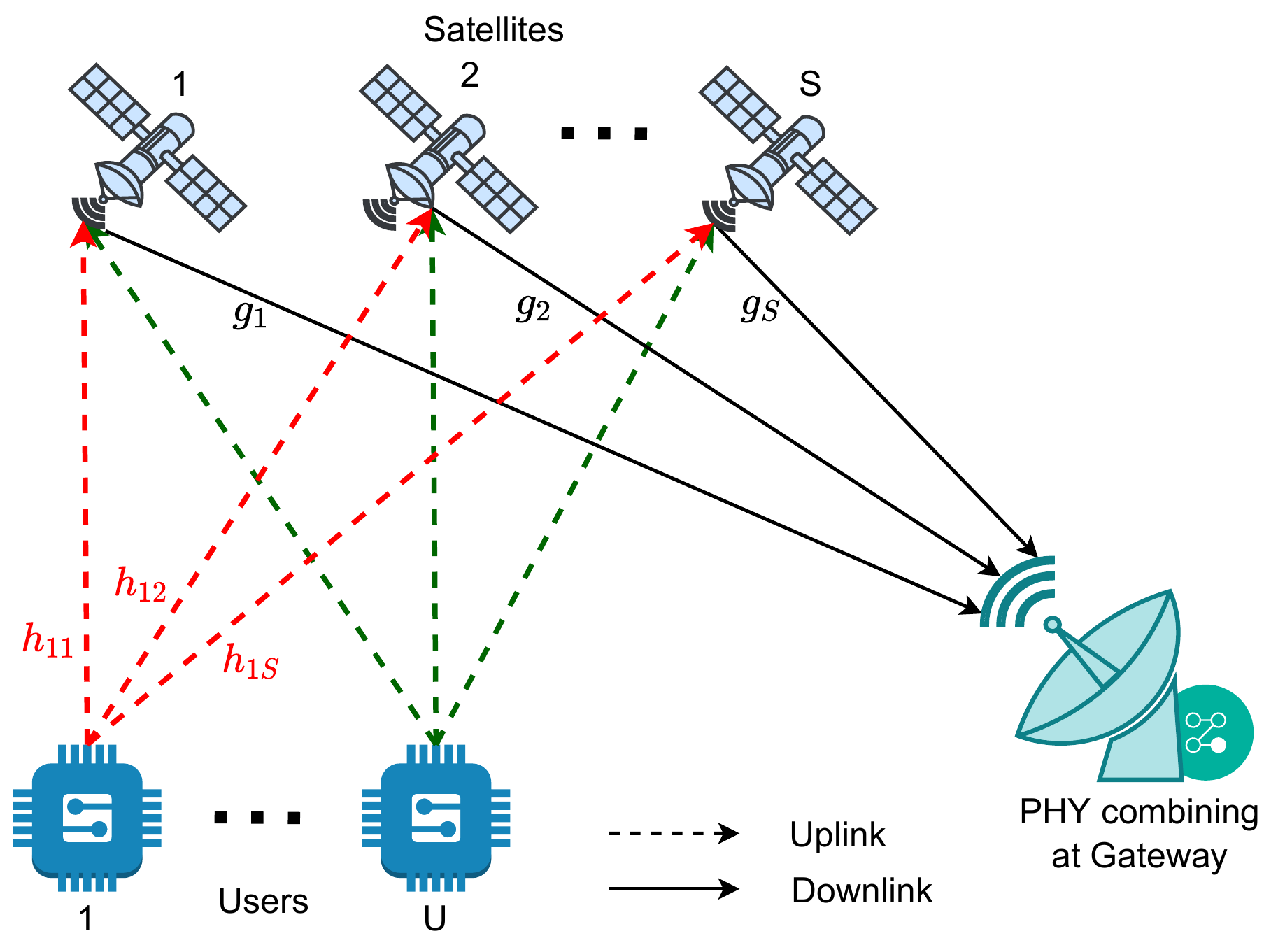}
\caption{Schematic diagram of the proposed topology with $U$ \ac{IoT} users broadcasting their sensed information to $S$ visible satellites simultaneously (different colours are used to indicate transmissions from each user). All the satellites forward the signals to the \ac{GS} using fixed-gain \ac{AF} relaying.}
\label{systemmodel}
\end{figure}
\begin{figure*}
\newcounter{mycnt0}
\setcounter{mycnt0}{\value{equation}}
\setcounter{equation}{5}
\begin{align}
    \gamma_{us} &= \frac{ r_{\text{min}}^{-\alpha}\, r_{us}^{-\alpha}\, G_s\, H_{us}}{ r_{\text{min}}^{-\alpha}\, G_s\, \left(\sum\limits_{\substack{i=1 \\ i\neq u}}^U r_{is}^{-\alpha} H_{is} + \sum\limits_{u=1}^U \eta_u r_{us}^{-\alpha} \sigma^2_{e_{us}} + 1\right) + \eta_s \sigma^2_{e_s} r_{\text{min}}^{-\alpha} \left(\sum\limits_{u=1}^U r_{us}^{-\alpha} \left(H_{us} + \eta_u \sigma^2_{e_{us}}\right) + 1\right)+ \dfrac{P_s}{\beta_{\text{AF}}^2\,\sigma_n^2}}, \nonumber\\[4pt]
    &= \frac{ r_{us}^{-\alpha}\, G_s\, H_{us}}{ G_s\, \left(\sum\limits_{\substack{i=1 \\ i\neq u}}^U r_{is}^{-\alpha} H_{is} + \sum\limits_{u=1}^U \eta_u r_{us}^{-\alpha} \sigma^2_{e_{us}} + 1\right) + \eta_s \sigma^2_{e_s} \left(\sum\limits_{u=1}^U r_{us}^{-\alpha} \left(H_{us} + \eta_u \sigma^2_{e_{us}}\right) + 1\right)+ \widehat{C}},
    \label{sinr_general}
\end{align}
\setcounter{equation}{\value{mycnt0}}
\hrule
\end{figure*}
As shown in Fig. \ref{systemmodel}, a direct access topology based on a mega-\ac{LEO} constellation is explored, where $U$ \ac{IoT} users communicate their sensed information to a \ac{GS} via $S$ satellites among all the $K_{\text{vis}}$ visible-satellites. The \ac{IoT} users are assumed to broadcast their information simultaneously using shared resources at the start of every slot as per the slotted-ALOHA scheme similar to the case shown in \cite{slotted_NOMA}. Keeping in mind the low complexity of \ac{IoT} users and design for a common application, it is assumed that all the users transmit at equal power. The visible satellites amplify and re-transmit the received information to the \ac{GS}. The \ac{GS} decodes the information of all the users after coherently combining the signals received from all satellites. Thus, end-to-end communication takes place in two phases. In the first phase, all the \ac{IoT} users who have sensed information broadcast their signal to all the satellites in the visible range. The signal received at the $s^\text{th}$ satellite can be written as
\begin{equation}
    y_s = \sum_{u = 1}^U \sqrt{P_u \mathcal{G}_u \mathcal{G}_s(\varphi_{us}) (\lambda / 4 \pi r_{us})^{\alpha}} (\hat{h}_{us}+e_{us}) x_u + n_s,
\end{equation}
where $P_u$ is the transmit power of the $u^\text{th}$ \ac{IoT} user, $r_{us}$ is the distance between the $u^\text{th}$ user and the $s^\text{th}$ satellite, $\alpha$ is the path loss exponent, $x_u$ is the unit energy information signal, $n_s$ is the \ac{AWGN} with zero mean and variance $\sigma_n^2$ at the satellite receiver, $\hat{h}_{us}$ is the \ac{SR} distributed imperfect estimate of the channel coefficient and $e_{us}$ is the estimation error. Similar to the approach followed in \cite{iCSI1,iCSI2,iCSI3,iCSI4}, the estimation error $e$ is considered to be distributed as $\mathcal{CN}(0,\sigma^2_e)$ with an SNR dependent variance $\sigma^2_e = \phi \, \eta^{-\chi}$, where $\phi$ and $\chi$ are deterministic constants. Here, a perfect CSI can be obtained if $\chi \rightarrow \infty$ for $\eta > 0$. The transmit and the receive antenna gains at the user and the satellite are denoted as $\mathcal{G}_u$ and $\mathcal{G}_s{(\varphi_{us})}$ respectively, where $\varphi_{us}$ is the angle between the $u^\text{th}$ user location and the beam center with respect to the $s^\text{th}$ satellite.

The \ac{SR} model best characterizes the channels which experience \ac{LoS} shadowing and small-scale fading \cite{sr}. It is a generalized form of a Rician fading model where the \ac{LoS} component is assumed to undergo Nakagami-$m$ fading. The \ac{SR} model is known to fit best the experimental data in the case of characterizing satellite models \cite{sr}. For any \ac{SR} random variable $h_i$, the \ac{PDF} and \ac{CDF} of $H_i = \eta_i |h_i|^2$,  are given, respectively in \cite{overlay} by
\vspace{-0.1cm}
\begin{align}
    f_{H_{i}}(x) =\, &\alpha_i \sum_{\kappa = 0}^{m_i-1} \frac{\zeta(\kappa)}{\eta_i^{\kappa + 1}} x^{\kappa} e^{-\left(\frac{\beta_i - \delta_i}{\eta_i} \right)x},
\label{pdf_sr} \\
F_{H_{i}}(x) =\,& 1 - \alpha_i \sum_{\kappa = 0}^{m_i-1} \frac{\zeta(\kappa)}{\eta_i^{\kappa + 1}} \sum_{p=0}^{\kappa}\frac{\kappa!}{p!} \left(\frac{\beta_i - \delta_i}{\eta_i} \right)^{-(\kappa + 1 - p)}\nonumber \\
& \times x^p  e^{-\left(\frac{\beta_i - \delta_i}{\eta_i} \right)x},
\label{cdf_sr}
\end{align}
where $\alpha_i = ((2b_i m_i)/(2b_i m_i +\Omega_i))^{m_i}/2b_i$, $\beta_i = 1/2b_i$, $\delta_i = \Omega_i/(2b_i)(2b_i m_i + \Omega_i)$ and $\zeta(\kappa) = (-1)^{\kappa} (1-m_i)_{\kappa} \delta_i^{\kappa}/(\kappa!)^2$ with $(\cdot)_\kappa$ being the Pochhammer symbol\cite{formula}. Here $2b_i$ denotes the average power of the multipath component, and $\Omega_i$ is the average power of the \ac{LoS} component. The \ac{LEO} satellites also observe Doppler shifts, but it has not been considered here to keep the analysis simple. It is assumed that the Doppler can be compensated using known techniques \cite{doppler}.

In the second phase, the satellites employ \ac{AF} to send the received signals to the \ac{GS} using dedicated orthogonal resources without interference. This assumption considers the downlink between the satellites and \ac{GS} to be resource sufficient. Moreover, it keeps the analysis simpler and focused on the effect of uplink, which is limited by the transmit power of the \ac{IoT} users. Similar to the store-and-forward scheme adopted in \cite{lacuna}, it is considered that satellites offload their information when they are nearest to the \ac{GS} within a time range, i.e. $r_g = r_{\text{min}}$. The $S$ satellites among the $K_{\text{vis}}$ visible satellites which offload their information to the \ac{GS} in the defined time range are considered for \ac{MRC}. The received signal from the $s^\text{th}$ satellite at the \ac{GS} can be written as \vspace{0.3cm}
\begin{equation}
    z_s = \beta_{\text{AF}}\sqrt{\mathcal{G}_s \, \mathcal{G}_{\text{GS}}\, (\varphi_{s}) \, (\lambda /4 \pi r_{\text{min}})^{\alpha}} \, (\hat{g}_{s}+e_{s})\, y_s + w_s,
\vspace{0.2cm}
\end{equation}
where $\hat{g_s}$ is the estimated \ac{SR} channel coefficient between the $s^\text{th}$ satellite and \ac{GS}, $e_s$ is the estimation error distributed as $\mathcal{CN}(0,\sigma^2_{e_{s}})$, $\beta_{\text{AF}}$ is the \ac{AF} gain factor and $w_s$ is the additive white Gaussian noise with zero mean and variance $\sigma_w^2$ at the \ac{GS} receiver. The transmit and the receive antenna gains at the satellite and the GS are denoted as $\mathcal{G}_s$ and $\mathcal{G}_{\text{GS}}{(\varphi_{s})}$ where $\varphi_{s}$ is the angle between the $s^\text{th}$ satellite location and the beam center with respect to the GS. Ideally, the effect of the channel between the user and the satellite is equalized by the \ac{AF} gain factor \cite{ray}. In this paper, the received signal is scaled at the satellite by a fixed-gain factor $\beta_{\text{AF}}$, which is inversely proportional to the total received power and is defined as
\begin{align}
    &\beta_{\text{AF}} = \nonumber \\[4pt]
    &\sqrt{\frac{P_s}{\sum\limits_{u=1}^U P_u \mathcal{G}_u \mathcal{G}_s(\varphi_{us}) \left(\frac{\lambda}{4 \pi}\right)^{\alpha} \mathbb{E}\left[r_{us}^{-\alpha}\left(|\hat{h}_{us}|^2 + \sigma^2_{e_{us}}\right)\right] + \sigma_n^2}}.
    \label{afFactor}
\end{align}

The instantaneous end-to-end \ac{SINR} of the information signal from the $s^\text{th}$ satellite for the $u^\text{th}$ user at the \ac{GS} can be written as \eqref{sinr_general}, where $H_{us} = \eta_{u}\,|\hat{h}_{us}|^2$ is the instantaneous \ac{SNR} of a user-satellite link with $\eta_u = \frac{P_u}{\sigma_n^2}\, \mathcal{G}_u \, \mathcal{G}_s(\varphi_{us}) \left(\lambda/4 \pi\right)^\alpha$, $G_{s} = \eta_{s}\,|\hat{g_s}|^2$ is the instantaneous \ac{SNR} of a satellite-\ac{GS} link with $\eta_s = \frac{P_s}{\sigma_w^2}\, \mathcal{G}_s \, \mathcal{G}_{\text{GS}}(\varphi_{s}) \left(\lambda/4 \pi\right)^\alpha$ and $\widehat{C} = \frac{P_s}{r_{\text{min}}^{-\alpha} \, \sigma_n^2 \, \beta_{\text{AF}}^2}$. Using \eqref{afFactor}, we can further simplify $\widehat{C}$ as
\vspace{0.2cm}
\begin{align}
    \setcounter{equation}{6}
    \widehat{C} &=\frac{1}{r_{\text{min}}^{-\alpha}}\left\{1+\sum\limits_{u=1}^U \left(\mathbb{E}\,[r_{us}^{-\alpha}H_{us}] + \eta_u \, \sigma^2_{e_{us}} \mathbb{E}[r_{us}^{-\alpha}]\right)\right\}.
\end{align}

Since the \ac{GS} combines the signals from all the visible satellites using \ac{MRC}, the end-to-end \ac{SINR} of the combined signal for the $u^\text{th}$ user at the \ac{GS} is given by
\begin{equation}
    \label{mrc}
    \gamma_u = \sum_{s=1}^S \gamma_{us}.
\end{equation}
\begin{figure*}[t!]
\centering
\subfloat[]{\includegraphics[width=2.4in]{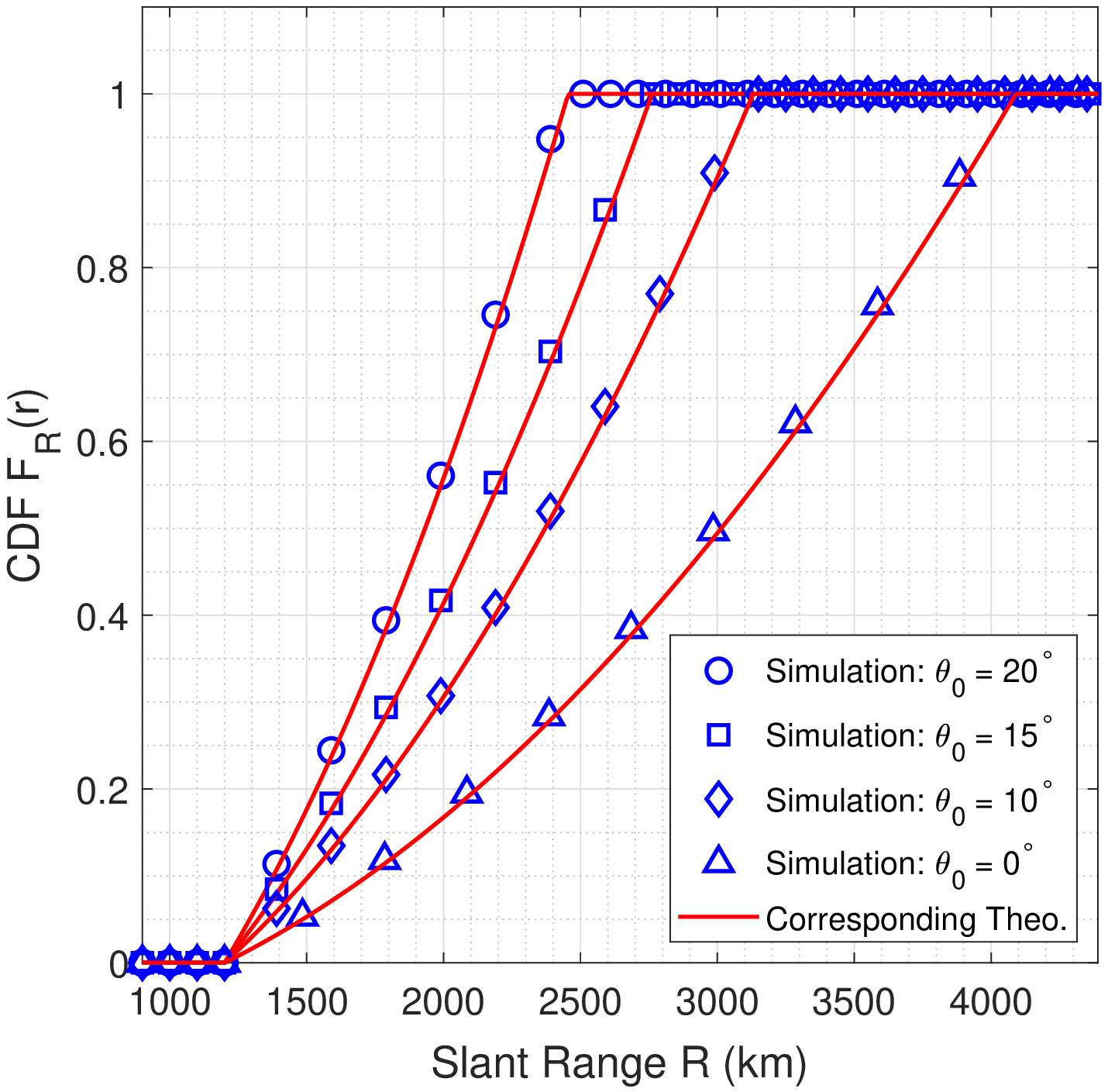}\label{DistancecDFEl}}
\hfil
\subfloat[]{\includegraphics[width=2.46in]{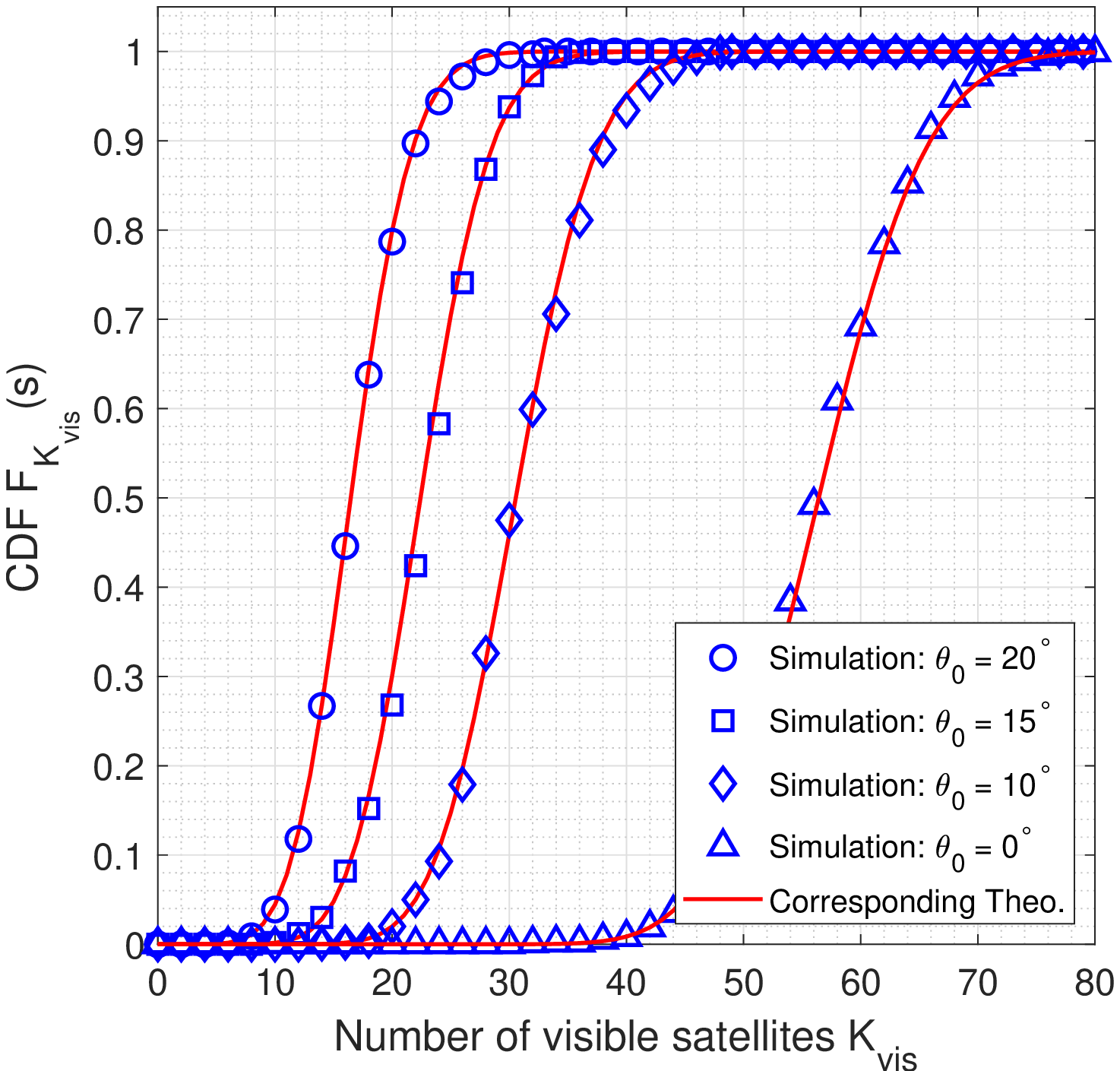}\label{numVisibleCDFEl}}
\caption{(a) CDF $F_R(r)$ of the distance between a user and a visible satellite in the constellation at 1200 km altitude. (b) CDF $F_{K_{\text{vis}}}(s)$ of the number of satellites visible to any user in a constellation of 720 satellites at 1200 km altitude.}
\vspace{0.1cm}
\end{figure*}
In this paper, two decoding schemes are compared to analyse the performance of the proposed topology:

\begin{itemize}
    \item \textit{Capture model (CM)}: The \ac{GS} is assumed to perfectly decode the information of the desired user out of many interfering signals if its \ac{SINR} is higher than a threshold. This type of decoding is similar to the capture effect used in LoRa \cite{capture}.
    \item \textit{Successive interference cancellation (SIC)}: The \ac{GS} decodes the information of the intended user by successively removing the information of other users in the order of their \ac{SINR}s \cite{nomatutorial}. The user with the highest \ac{SINR} is decoded first, and its reconstructed signal is subtracted from the received superimposed signal to decode the remaining users. However, even after removing the interference, there may still be some residual error remaining due to noise or imperfect decoding. The user with \ac{SINR} less than the threshold and subsequent users in the order are considered non-decodable and contribute to the outage.
\end{itemize}
%
%
\vspace{0.2cm}
\section{Slant distance and the number of visible satellites}\label{R_Kvis}
Since the satellites are considered to be distributed on a spherical surface following a \ac{BPP}, the distance between a user and a satellite is random. Moreover, the total number of visible satellites is also random and depends upon $\theta_0$ and the total number of satellites in the constellation $K$. Statistical characteristics of the slant distance and number of visible satellites are derived in this section.
\subsection{Statistical characteristics of the distance between the user and the satellite}\label{PDF_CDF_Range}
The \ac{CDF} of the distance $R$ between a user and visible satellites in the constellation is given by
\begin{equation}\label{cdfR}
 F_R(r) = 
  \begin{cases} 
   0, & r < r_{\textup{min}} \\[2pt]
   \dfrac{r^2 - r_{\textup{min}}^2}{r_{\textup{max}}(\theta_0)^2 - r_{\textup{min}}^2},       & r_{\textup{min}} \le r \le r_{\textup{max}}(\theta_0) \\
   & 0 \le \theta_0 < 90^{\circ}\\[2pt]
   1, & r > r_{\textup{max}},
  \end{cases}
\end{equation}
and the corresponding PDF is given by 
\begin{equation}\label{pdfR}
    f_R(r) =
    \begin{cases} 
        \dfrac{2\,r}{r_{\textup{max}}(\theta_0)^2 - r_{\textup{min}}^2}, & r_{\textup{min}} \le r \le r_{\textup{max}}(\theta_0) \\
        & 0 \le \theta_0 < 90^{\circ}\\[2pt]
        0, &\text{otherwise},
    \end{cases}
\end{equation}
where $r_{\textup{min}} = d$ is the orbital altitude and  $r_{\textup{max}}(\theta_0) = \sqrt{(r_e \sin \theta_0)^2 + (r_e+r_{\text{min}})^2 - r_e^2} -r_e \sin\theta_0$ is the maximum distance observed at mask elevation angle $\theta_0$. The proofs for \eqref{cdfR} and \eqref{pdfR} are provided in Appendix \ref{proof:range}. The effect of $\theta_0$ on the range can be inferred from Fig. \ref{DistancecDFEl}. It can be observed that the derived expressions match the simulation results. The slant range for which the \ac{CDF} reaches 1 corresponds to the maximum possible range for a user. As the mask angle increases, the maximum possible range decreases. However, it can be observed that the maximum range decreases rapidly with an increase in the mask angle from $0^\circ$ to $10^\circ$ when compared to $10^\circ$ to $20^\circ$. It can be attributed to the fact that, $r_{\text{max}}$ changes non-linearly with change in $\theta_0$ as shown in \eqref{rMax}.

\vspace{0.2cm}
\textbf{Remark:} For the special case of $\theta_0 = 0^\circ$, $r_{\text{max}}(0^{\circ}) = \sqrt{r_{\text{min}}^2 + 2 r_e r_{\text{min}}}$. Hence for $r_{\textup{min}} \le r \le r_{\textup{max}}(0^{\circ})$, \eqref{cdfR} and \eqref{pdfR} can be simplified as
\begin{align}
    F_{R_0}(r) &= \frac{r^2 - r_{\text{min}}^2}{2\,r_e\,r_{\text{min}}}, \label{cdfR_El0}\\
    f_{R_0}(r) &= \frac{r}{r_e\,r_{\text{min}}}, \label{pdfR_El0}
\end{align}
where $R_0$ denotes the random variable for $R$ at $\theta_0 = 0^\circ$. The expressions in \eqref{cdfR_El0} and \eqref{pdfR_El0} match with the expressions given for the characteristics of the distance in \cite{stochastic2}, where they are derived for $\theta_0=0^\circ$ only. The simplified expressions shown above are used for scenarios where high-rise structures like mountains and buildings do not mask satellite visibility.
\begin{figure*}
\newcounter{mycnt1}
\setcounter{mycnt1}{\value{equation}}
\setcounter{equation}{17}
\begin{align}
    F_{\gamma_{us} \,|\, r_{us}}(x) &= 1 - \sum\limits_{k_1=0}^{m_{us}-1}\sum\limits_{p=0}^{k_1} \alpha_{us}\, \frac{k_1!\,\zeta(k_1)}{p!\,\eta_{us}^{k_1+1}} A_1^{-(k_1+1-p)} \sum\limits_{k_2=0}^{m_s-1} \alpha_s \frac{\zeta(k_2)}{\eta_{s}^{k_2+1}} \sum\limits_{z=0}^{p} \binom{p}{z} \left(\frac{x\,a}{r_{us}^{-\alpha}}\right)^{p-z} \exp \left[-A_1 \left(\frac{x\,a}{r_{us}^{-\alpha}}\right)\right]\left(\frac{x\,C}{r_{us}^{-\alpha}}\right)^z \nonumber\\
    & \quad \times \int\limits_{g=0}^{\infty} g^{-z+k_2} \, \exp\left[-A_1\left(\frac{x\,C}{r_{us}^{-\alpha}\,g}\right) - A_2\,g\right] \, dg, \label{cdf_cm_2}\\
    F_{\gamma_{us} \,|\, r_{us}}(x)&= 1 - \sum\limits_{k_1=0}^{m_{us}-1}\sum\limits_{p=0}^{k_1} \alpha_{us}\, \frac{k_1!\,\zeta(k_1)}{p!\,\eta_{us}^{k_1+1}} A_1^{-(k_1+1-p)} \sum\limits_{k_2=0}^{m_s-1} \alpha_s \frac{\zeta(k_2)}{\eta_{s}^{k_2+1}} \sum\limits_{z=0}^{p} \binom{p}{z} \left(\frac{x\,a}{r_{us}^{-\alpha}}\right)^{p-z} \exp \left[-A_1 \left(\frac{x\,a}{r_{us}^{-\alpha}}\right)\right]\left(\frac{x\,C}{r_{us}^{-\alpha}}\right)^z \nonumber\\
    & \quad \times 2\, \left(\frac{A_1\,x\,C}{A_2\, r_{us}^{-\alpha}}\right)^{(1-z+k_2)/2}\, K_{1-z+k_2}\left(2\,\sqrt{\frac{A_1 A_2\, x\, C}{r_{us}^{-\alpha}}}\right), \label{cdf_cm_3}
\end{align}
\hrule
\setcounter{equation}{\value{mycnt1}}
\vspace{0.2cm}
\end{figure*}
\subsection{Statistical characteristics of the number of visible satellites}\label{PDF_CDF_S}
A satellite is visible to a user only if its elevation angle exceeds the minimum required elevation $\theta_0$, also called the mask angle. For a given mask elevation angle $\theta_0$, the number of visible satellites $K_{\text{vis}}$ to any user is a binomial random variable with success probability
\begin{equation}\label{trueP}
\mathcal{P} = \dfrac{r_{\textup{max}}(\theta_0)^2 - r_{\textup{min}}^2}{4\, r_e\, (r_e + r_{\textup{min}})},
\end{equation}
where $r_{\textup{max}}(\theta_0)$ is the distance observed at $\theta_0$. The proof for \eqref{trueP} is provided in Appendix \ref{proof:numVisible}. The effect of mask angle on the number of visible satellites can be inferred from Fig. \ref{numVisibleCDFEl}. It can be observed that the derived expressions match the simulation results. The number of visible satellites for which \ac{CDF} equals 1 denotes the maximum possible number of satellites which can be visible to a user. As the mask angle increases, the surface area of the cap region shown in Fig. \ref{sm} decreases, and so does the maximum number of visible satellites. It can be observed that the maximum possible number of visible satellites decreases rapidly from $0^\circ$ to $10^\circ$ when compared to $10^\circ$ to $20^\circ$. It can be understood using \eqref{rMax} and \eqref{eqn:SVis_2}, since $A_{\text{vis}}$ decreases non-linearly with an increase in $\theta_e$.\looseness=-1

\vspace{0.2cm}
\textbf{Remark:} With $\mathcal{P}_0$ denoting the success probability for the special case of $\theta_0 = 0^\circ$, \eqref{trueP} can be simplified as
\begin{align}
    \mathcal{P}_0 = \frac{r_{\text{min}}}{2\,(r_e + r_{\text{min}})}.
\end{align}
%
\section{Outage Probability Analysis}\label{OPDerivation}
The outage probability of a particular user is defined as
\begin{align}
    P_{\text{out}}(\mathcal{R}) & \triangleq \p\left[\frac{B}{2}\log_2(1+\text{SINR}) \leq \mathcal{R}\right]\nonumber\\
    &= \p\left[\text{SINR} \le \gt \right],
\end{align}
where $\mathcal{R}$ is the target rate, $B$ is the bandwidth,  $\gamma_{\text{th}} \triangleq  2^{2\mathcal{R}/B} - 1$ is the threshold and \ac{SINR} needs to be calculated for \ac{CM} and \ac{SIC} schemes separately.
\subsection{OP for CM based decoding}\label{OPCapture}
In \ac{CM} based decoding, a user is decoded in the presence of interference from all other users. Hence the \ac{OP} of a user at the \ac{GS} can be written as
\begin{align}
    \label{opCM}
    P_{\text{out}}(\mathcal{R}) &= \left(\p\left[\gamma_u \le \gt \right] \times \p[K_{\text{vis}} \ge S]\right) + \p[K_{\text{vis}} < S]  \nonumber\\[4pt]
    &=\left(F_{\gamma_{u}}(\gt) \times \p[K_{\text{vis}} \ge S]\right) + \p[K_{\text{vis}} < S],
\end{align}
where $\gamma_u = \sum_{s=1}^S \gamma_{us}$ and $\p[K_{\text{vis}} \ge S] = 1-\sum_{j=0}^{S-1} {K \choose j}\, \mathcal{P}^j\, (1-\mathcal{P})^{K-j} $. The following three-step approach has been followed to find the exact expression for \eqref{opCM}.
\begin{enumerate}
    \item Finding the \ac{CDF} of $\gamma_{us}$ conditioned on $r_{us}$ for a single satellite scenario.
    \item Finding the \ac{MGF} of $\gamma_{us}$ and $\gamma_{u}$ for extending the analysis to the multi-satellite scenario.
    \item Finding the \ac{CDF} of $\gamma_{u}$ and consequently the \ac{OP} in multi-satellite scenario.
\end{enumerate}

\vspace{0.2cm}
\textit{Step 1:} Using the theorem of transformation of random variables, the \ac{CDF} of $\gamma_{us}$ conditioned on the distance between the user and the satellite can be found as
\begin{align}
   &F_{\gamma_{us} \,|\, r_{us}}(x) 
   = \p[\gamma_{us} \le x\,| \,r_{us}] \nonumber\\[4pt]
   &\stackrel{(\text{i})}{\approx} \p \left[ \left.\frac{ r_{us}^{-\alpha}\, G_s\, H_{us}}{ G_s\, \left((U-1)\bar{I} + U\, \overline{E}_1 + 1\right) + \overline{E}_2 + \widehat{C}} \le x \,\right|\, r_{us}\right]\nonumber\\[4pt]
   &\stackrel{(\text{ii})}{=}\int\limits_{g=0}^{\infty}F_{H_{us}}\left(\frac{x\,a}{r_{us}^{-\alpha}} + \frac{x\,C}{r_{us}^{-\alpha}\,g}\right)\, f_{G_s}(g)\,\, dg, \label{cdf_cm_1}
\end{align}
where in (i), interference for a user and error terms due to \ac{CSI} mismatch are approximated as $\bar{I} = \mathbb{E}[r_{is}^{-\alpha}\, H_{is}]$, $\overline{E}_1 = \eta_u \, \sigma^2_{e_{us}} \, \mathbb{E}[r_{is}^{-\alpha}]$ and $\overline{E}_2 = \eta_s \, \sigma^2_{e_s} \,U\, (\bar{I} + \overline{E}_1)$ for mathematical tractability and in (ii), $a = (U-1)\, \bar{I} + U\overline{E}_1 + 1$ and $C = \overline{E}_2 + \widehat{C}$ for convenience. The expressions for the terms $\mathbb{E}[r_{is}^{-\alpha}]$ and $\mathbb{E}[r_{is}^{-\alpha}\, H_{is}]$ are derived in Appendix \ref{proof:ExpectationRH}. Using \eqref{pdf_sr}, \eqref{cdf_sr}, with the knowledge of binomial expansion and interchanging the order of summation and integration, \eqref{cdf_cm_1} can be simplified as \eqref{cdf_cm_2}, where $A_1 = \frac{\beta_{us} - \delta_{us}}{\eta_{us}}$ for uplink and $A_2 = \frac{\beta_{s} - \delta_{s}}{\eta_{s}}$ for downlink. The integral expression in \eqref{cdf_cm_2} can be solved using \cite[Eq. 3.471.9]{formula} to get the closed-form expression for $F_{\gamma_{us} \,|\, r_{us}}(x)$ as shown in \eqref{cdf_cm_3}, where $K_v(\cdot)$ is the $v^{\text{th}}$ order modified Bessel function of second kind.
\begin{figure*}
\setcounter{equation}{21}
\begin{align}
    M_{\gamma_{us} \,|\, r_{us}}(-t) &= 1 - t\, \sum\limits_{k_1=0}^{m_{us}-1}\sum\limits_{p=0}^{k_1} \alpha_{us}\, \frac{k_1!\,\zeta(k_1)}{p!\,\eta_{us}^{k_1+1}}\, A_1^{-(k_1+1-p)} \sum\limits_{k_2=0}^{m_s-1} \alpha_s\, \frac{\zeta(k_2)}{\eta_{s}^{k_2+1}}\, \sum\limits_{z=0}^{p} {p \choose z}\, (a\,r_{us}^{\alpha})^{p-z}\, (C\,r_{us}^{\alpha})^{(1+z+k_2)/2}\nonumber\\
    & \quad \times \left(\frac{A_1}{A_2}\right)^{(1-z+k_2)/2}\, \epsilon^{-\mu}\,\exp\left[\frac{\Lambda^2}{2\,\epsilon}\right]\, \frac{\Gamma(\mu + \nu + \frac{1}{2})\Gamma(\mu - \nu + \frac{1}{2})}{\Lambda}\, W_{-\mu,\nu}\left(\frac{\Lambda^2}{\epsilon}\right),\label{mgf_2}
\end{align}
\hrule
\setcounter{equation}{\value{mycnt1}}
\end{figure*}

\vspace{0.2cm}
\textit{Step 2:} For any random variable $X$, with \ac{MGF} $M_X(t)$ and $\mathcal{L}\{\cdot\}$ denoting the Laplace transform operator, we can write
\setcounter{equation}{19}
\begin{align}
    \mathcal{L}\{f_X(x)\} &= M_X(-t), \nonumber\\
    \mathcal{L}\{F_X(x)\} &= \frac{M_X(-t)}{t},\label{mgfPro_2}
\end{align}
where \eqref{mgfPro_2} follows from the integral property of Laplace transform. Therefore, flipped \ac{MGF}, $M_X(-t)$ is required to obtain \ac{CDF} by applying the inverse Laplace transform on \eqref{mgfPro_2}. The flipped \ac{MGF} (referred simply as \ac{MGF} hereafter) of $\gamma_{us}$ conditioned on $r_{us}$ can be derived using the definition of the Laplace transform as
\begin{align}\label{mgf_1}
    M_{\gamma_{us} \,|\, r_{us}}(-t) = 1 - t\int\limits_{x=0}^{\infty} e^{-tx} \left(1-F_{\gamma_{us} \,|\, r_{us}}(x)\right) \, dx.
\end{align}
Using \eqref{cdf_cm_3} and \cite[Eq. 6.643.3]{formula}, the integral in \eqref{mgf_1} can be solved to arrive at the closed-form expression for $M_{\gamma_{us} \,|\, r_{us}}(-t)$ as shown in \eqref{mgf_2}, where
\begin{align}
    \mu &= p + 1 + \frac{k_2-z}{2}, \nonumber\\
    \epsilon &= A_1\, a\, r_{us}^{\alpha} + t ,\nonumber\\
    \Lambda &= \sqrt{A_1\,A_2\,C\,r_{us}^{\alpha}}\,, \nonumber\\
    \nu &= \frac{1-z+k_2}{2} ,\nonumber
\end{align}
and $\Gamma(\cdot)$, $W_{\mu,v}(\cdot)$ are the Gamma and Whittaker functions, respectively.\looseness=-1

The \ac{MGF} of $\gamma_{us}$ can be calculated by averaging over $r_{us}$ using \eqref{pdfR} and \eqref{mgf_2} as
\setcounter{equation}{22}
\begin{align}
    M_{\gamma_{us}}(-t) &= \int\limits_{r_{\text{min}}}^{r_{\text{max}}} M_{\gamma_{us} \,|\, r_{us}}\left(-t\right) f_R(r_{us}) \, dr_{us} \nonumber\\
    &= \frac{2}{r_{\textup{max}}^2 - r_{\textup{min}}^2} \int\limits_{r_{\text{min}}}^{r_{\text{max}}}\, r_{us}\, M_{\gamma_{us} \,|\, r_{us}}\left(-t\right) \, dr_{us}.\label{mgf_3}
\end{align}
The integral in \eqref{mgf_3} can be efficiently evaluated using numerical techniques as discussed in Appendix \ref{App:int}.

The \ac{MRC} is implemented at the \ac{GS} on the signal with end-to-end \ac{SINR} as defined in \eqref{mrc}. Since all the $S$ satellite-\ac{GS} links are independent, the \ac{MGF} of the end-to-end \ac{SINR} can be written as
\begin{align}
    &M_{\gamma_u}(-t) = \prod_{s=1}^S M_{\gamma_{us}}(-t).
\end{align}

\textit{Step 3:} Using \eqref{mgfPro_2}, the \ac{CDF} of $\gamma_u$ as be obtained as
\begin{align}
     F_{\gamma_{u}}(x) = \mathcal{L}^{-1}\left\{\frac{M_{\gamma_u}(-t)}{t}\right\}\left(x\right).\label{cdfCM_CF}
\end{align}
The inverse Laplace transform in \eqref{cdfCM_CF} can be efficiently calculated using the numerical technique presented in \cite{laplaceInv} as
\begin{align}
    F_{\gamma_{u}}(x) = &\frac{2^{-Q}\, e^{D/2}}{x} \sum\limits_{q=0}^{Q} {Q \choose q} \nonumber\\
    &\sum\limits_{n=0}^{N+q} \frac{(-1)^n}{\Delta_n} \Re\left\{\dfrac{M_{\gamma_u}\left(-\dfrac{D + 2\pi j n}{2\,x}\right)}{\dfrac{D + 2\pi j n}{2\,x}}\right\} \nonumber \\
    & + E(D,Q,N),
    \label{cdfCM_CF_1}
\end{align}
where
\begin{align}
    \Delta_n = &
    \begin{cases}
        2, &n = 0\\
        1, &n = 1,2,\cdots,N
    \end{cases}\nonumber
\end{align}
and
\begin{align}
    E(D,Q,N) = & \frac{e^{-D}}{1-e^{-D}} + \frac{2^{-Q}\, e^{D/2}}{x} \sum\limits_{q=0}^{Q} (-1)^{N+1+q} {Q \choose q} \nonumber\\
    &\Re\left\{\dfrac{M_{\gamma_u}\left(-\dfrac{D + 2\pi j(N+q+1)}{2\,x}\right)}{\dfrac{D + 2\pi j(N+q+1)}{2\,x}}\right\}.\nonumber
\end{align}
The values of $D$, $Q$ and $N$ are selected to keep the discretization and truncation errors negligible. Thus, using \eqref{trueP} and \eqref{cdfCM_CF_1} in \eqref{opCM}, completes the derivation of \ac{OP} for \ac{CM}-based decoding.\looseness=-1
\begin{figure*}
\newcounter{mycnt3}
\setcounter{mycnt3}{\value{equation}}
\setcounter{equation}{33}
    \begin{align}
        \label{cdfHBar_1}
        F_{\widetilde{H}_{us}}(z) &= \int\limits_{r_{\text{min}}}^{r_{\text{max}}} F_{H_{us}}\left(\frac{z}{r_{us}^{-\alpha}}\right) f_{R}(r_{us}) \,dr_{us} = 1 - \frac{2\,\alpha_{us}}{r_{\textup{max}}^2 - r_{\textup{min}}^2} \sum\limits_{k=0}^{m_{us}-1}\sum\limits_{p=0}^{k} \,\frac{k!\,\zeta(k)}{p!\,\eta_{us}^{k+1}} A^{-(k+1-p)} \int\limits_{r_{\text{min}}}^{r_{\text{max}}} (z\,r_{us}^{\alpha})^p \, \exp[{-Azr_{us}^{\alpha}}] \,r_{us}\,dr_{us}, \\
        F_{\widetilde{H}_{us}}(z) &= 1 - \frac{2}{r_{\textup{max}}^2 - r_{\textup{min}}^2} \sum\limits_{k=0}^{m_{us}-1}\sum\limits_{p=0}^{k} \alpha_{us}\, \frac{k!\,\zeta(k)}{p!\,\eta_{us}^{k+1}}\, A^{-(k+1-p)}\, \left(\frac{\gamma(V,\rho\,r_{\text{max}}^\alpha) - \gamma(V,\rho\,r_{\text{min}}^\alpha)}{\alpha\,\rho^V}\right)\,z^p.
        \label{cdfHBar_2}
    \end{align}
    \hrule
    \vspace{0.2cm}
    \setcounter{equation}{36}
    \begin{align}
        \label{mgfSIC_2}
        M_{\gamma_{us}^{(1)}|G_s}(-t) &= 1 - \frac{2\,t}{r_{\textup{max}}^2 - r_{\textup{min}}^2} \sum\limits_{k=0}^{m_{us}-1}\sum\limits_{p=0}^{k} \frac{\alpha_{us}}{\alpha}\, \frac{k!\,\zeta(k)}{p!\,\eta_{us}^{k+1}} A^{-(k-p+V+1)}\,\left(a+\frac{C}{g_s}\right)^{p-V} \nonumber\\
        &\times \int\limits_{x=0}^{\infty} e^{-tx}\,x^{p-V}\,\left\{\gamma\left(V,A\,r_{\text{max}}^\alpha\,\left(a+\frac{C}{g_s}\right)x\right) - \gamma\left(V,A\,r_{\text{min}}^\alpha\,\left(a+\frac{C}{g_s}\right)x\right)\right\}\, dx .\\
        \label{mgfSIC_3}
        M_{\gamma_{us}^{(1)}|G_s}(-t) &=1 - \frac{2\,t}{r_{\textup{max}}^2 - r_{\textup{min}}^2} \sum\limits_{k=0}^{m_{us}-1}\sum\limits_{p=0}^{k} \frac{\alpha_{us}}{\alpha}\, \frac{k!\,\zeta(k)}{p!\,\eta_{us}^{k+1}} A^{-(k-p+V+1)}\,\left(a+\frac{C}{g_s}\right)^{p-V} \, \frac{\Gamma(p+1)}{V} \nonumber\\
        & \times \left\{ \frac{\left[\upsilon_1(g_s)\right]^V }{\left[\upsilon_1(g_s) + t\right]^{p+1}}\,\,_2F_1\left(1,p+1;V+1;\frac{\upsilon_1(g_s)}{\upsilon_1(g_s)+t}\right) \,-\, \frac{\left[\upsilon_2(g_s)\right]^V }{\left[\upsilon_2(g_s) + t\right]^{p+1}}\,\,_2F_1\left(1,p+1;V+1;\frac{\upsilon_2(g_s)}{\upsilon_2(g_s)+t}\right) \right\},
    \end{align}
    \hrule
    \setcounter{equation}{\value{mycnt3}}
\end{figure*}
%
\subsection{OP for SIC based decoding}\label{OPSIC}
\ac{SIC} is an ordering-based scheme where the \ac{GS} decodes the information of users in the order of their end-to-end \ac{SINR}s. The residual error due to imperfect interference cancellation is considered to be distributed as $\mathcal{CN}(0,\xi)$ where $\xi$ represents the power of the residual error. For the ease of understanding, $l \in [1,U]$ is used to denote the order/iteration of \ac{SIC} decoding and $\mathbf{D}[\,l\,]$ is used to denote the set of indexes for all decoded users till the $l^{\text{th}}$ iteration. Additionally, $\mathbf{D}\{l\}$ is used to denote the index of the user decoded at the $l^{\text{th}}$ iteration. Therefore end-to-end \ac{SINR} of the signal from $u^{\text{th}}$ user received via $s^{\text{th}}$ satellite in the $l^{\text{th}}$ iteration of \ac{SIC} decoding can be written as
\begin{align}
    \gamma_{us}^{(l)} = \frac{ r_{us}^{-\alpha}\, G_s\, H_{us}}{ G_s\, \left((U-l)\bar{I} + U\overline{E}_1 + (l-1)\overline{E}_3 + 1\right) + \overline{E}_2 + \widehat{C}} \nonumber \\
    \forall \, u \notin \mathbf{D}[\,l-1\,],
\end{align}
where $\overline{E}_3 = \eta_u \, \xi \, \mathbb{E}[r_{is}^{-\alpha}]$. Also, the end-to-end \ac{SINR} of the \ac{MRC} combined signal of the $u^{\text{th}}$ user can be written as
\begin{align}
    \gamma_u^{(l)} = \sum_{s=1}^{S} \gamma_{us}^{(l)} \quad \forall \, u \notin \mathbf{D}[\,l-1\,].
\end{align}
At every iteration, the user with the highest \ac{SINR} is decoded such that
\begin{align}
    \gamma_{\mathbf{D}\{l\}}^{(l)} = 
    \begin{cases}
        \quad \ \max\limits_{u} \gamma_{u}^{(l)}, & l=1\\[8pt]
        \max\limits_{u, u \notin \mathbf{D}[\,l-1\,]} \gamma_{u}^{(l)}, & \gamma_{\mathbf{D}\{l-1\}}^{(l-1)} > \gt,\, l > 1 \\[8pt]
        \quad \ 0, & \text{otherwise},
    \end{cases}
\end{align}
where the set of indexes for all the decoded users is updated after every iteration as
\begin{align}
    &\mathbf{D}[\,l-1\,]= \nonumber\\
    & 
    \begin{cases}
        \O, &l=1 \\[8pt]
        \mathbf{D}[\,l-2\,] \cup \{\mathop{\arg\max}\limits_{\substack{u, u \notin \mathbf{D}[l-2]}} \gamma_u^{(l-1)}\}, & \gamma_{\mathbf{D}\{l-1\}}^{(l-1)} > \gt,l>1 \\[8pt]
        \mathbf{D}[l-2], & \gamma_{\mathbf{D}\{l-1\}}^{(l-1)} \leq \gt, l>1.
    \end{cases}
\end{align}
Therefore the \ac{OP} of $l^{\text{th}}$ user in case of \ac{SIC} can be written as
\setcounter{equation}{30}
\begin{align}
    \label{opSIC_1}
    &P_{\text{out}}^{(l)}(\mathcal{R}) =\nonumber\\[6pt]
    &
    \begin{cases}
        \left( \p [\max\limits_{u} \gamma_u^{(l)} \leq \gt]\, \p[K_{\text{vis}} \ge S]\right) + \p[K_{\text{vis}} < S], & l=1\\[10pt]
        \left\{\p [\gamma_{\mathbf{D}\{l\}}^{(l)} \leq \gt \,|\, \gamma_{\mathbf{D}\{l-1\}}^{(l-1)} > \gt]\, (1-P_{\text{out}}^{(l-1)}(\mathcal{R})) \right.\\
        \left. \ + \, P_{\text{out}}^{(l-1)}(\mathcal{R})\right\}\, \p[K_{\text{vis}} \ge S] + \p[K_{\text{vis}} < S], &l> 1,
    \end{cases}
\end{align}
or simply as
\begin{align}
    P_{\text{out}}^{(l)}(\mathcal{R}) = \left( F_{\gamma_{u}^{(l)}}(\gt) \times \p[K_{\text{vis}} \ge S]\right) + \p[K_{\text{vis}} < S].
    \label{opSIC_CF}
\end{align}

The exact expression of \eqref{opSIC_CF} for $l=1$ can be obtained. However, for $l>1$ it is not mathematically tractable since the distribution of $\gamma_{\mathbf{D}\{l\}}^{(l)}$ conditioned on $\gamma_{\mathbf{D}\{l-1\}}^{(l-1)} > \gt$ is difficult to obtain. Therefore, the exact expression for \ac{OP} in case of \ac{SIC} is obtained for the best user ($l=1$) only, which is a lower bound on the average \ac{OP} of the system and simulation results are presented for $l>1$. For the case of $l=1$, distribution for maximum of dependent random variables $\gamma_u^{(1)}$ is required. Hence the derivation is done using the following steps:
\begin{enumerate}
    \item Finding the \ac{CDF} of $\gamma_{us}^{(1)}$ conditioned on $G_s$ for single satellite scenario.
    \item Finding the \ac{MGF} of $\gamma_{us}^{(1)}$ and $\gamma_{u}^{(1)}$ conditioned on $G_s$ for extending the analysis to multi-satellite scenario.
    \item Finding the \ac{CDF} and consequently the \ac{OP} for $\max\limits_{u} \gamma_u^{(1)}$ averaged over all $G_s$.
\end{enumerate}

\textit{Step 1:} Using the theorem of transformation of random variables, the \ac{CDF} of $\gamma_{us}^{(1)}$ conditioned on $G_s$ can be written as:
\begin{align}
    F_{\gamma_{us}^{(1)}|G_s} &= \p\left[\left.\frac{ r_{us}^{-\alpha}\, G_s\, H_{us}}{ a \, G_s + C} \leq x \,\right|\,G_s\right] \nonumber\\
    &= \p \left[r_{us}^{-\alpha}\, H_{us} \leq a\,x + \left.\frac{C\,x}{G_s} \,\right|\,G_S\right] \nonumber\\
    &= F_{\widetilde{H}_{us}\mid G_s}\left(a\,x + \frac{C\,x}{g_s}\right),
\end{align}
where $a = \left((U-1)\bar{I} + U\overline{E}_1 + 1\right)$, $C = \overline{E}_2 + \widehat{C}$ and $\widetilde{H}_{us} = r_{us}^{-\alpha}\, H_{us}$. The \ac{CDF} $F_{\widetilde{H}_{us}}(z)$ can be written as \eqref{cdfHBar_1} where the integral can be solved using \cite[Eq. 3.381.8]{formula} to obtain \eqref{cdfHBar_2}. In \eqref{cdfHBar_2}, $A = \frac{\beta - \delta}{\eta}$, $V = \frac{\alpha\,p + 2}{\alpha},\  \rho = A\, z$ and $\gamma(\cdot,\cdot)$ is the lower incomplete Gamma function.

\textit{Step 2:} Similar to the approach followed in the derivation of \ac{CM} decoding, the \ac{MGF} of $\gamma_{us}^{(1)}$ conditioned on $G_s$ can be written as
\setcounter{equation}{35}
\begin{align}\label{mgfSIC_1}
    & M_{\gamma_{us}^{(1)}|G_s}(-t) = \nonumber\\
    & \quad \quad 1 - t\int\limits_{x=0}^{\infty} e^{-tx} \left(1-F_{\widetilde{H}_{us}\mid G_s}\left(ax + \frac{Cx}{g_s}\right)\right) dx.
\end{align}
Rearranging the terms, \eqref{mgfSIC_1} can be written as \eqref{mgfSIC_2}. The integral in \eqref{mgfSIC_2} can be solved using \cite[Eq. 6.455.2]{formula} to obtain \eqref{mgfSIC_3}, where
\setcounter{equation}{38}
\begin{align}
    \upsilon_1(g_s) = &A\,r_{\text{max}}^\alpha \,\left(a + \frac{C}{g_s}\right), \\
    \upsilon_2(g_s) = &A\,r_{\text{min}}^\alpha \,\left(a + \frac{C}{g_s}\right),
\end{align}
and $_2F_1(\cdot)$ is the Gauss hypergeometric function. Since all the $S$ satellites-\ac{GS} links are independent, the \ac{MGF} of $\gamma_u^{(1)}$ conditioned on $G_s$ can therefore be written as
\begin{align}
    &M_{\gamma_u^{(1)}|G_s}(-t) = \prod_{s=1}^S M_{\gamma_{us}^{(1)}|G_s}(-t).
\end{align}

\textit{Step 3:} Using \eqref{mgfPro_2} and averaging over all the $G_s$, the \ac{CDF} of $\max\limits_{u} \gamma_u^{(1)}$ can be derived as
\begin{align}
    F_{\gamma_{u}^{(1)}}(x) = &\int \limits_{g_1} \cdots \int \limits_{g_S} \left[\mathcal{L}^{-1}\left\{\frac{M_{\gamma_u^{(1)}|G_s}(-t)}{t}\right\}(x) \right]^U \nonumber\\
    &\times \left\{\prod_{s=1}^S f_{G_s}(g_s)\right\} \,dg_1 \cdots dg_S.
    \label{cdfSIC_CF}
\end{align}
The integral in \eqref{cdfSIC_CF} can be efficiently evaluated using numerical techniques similar to the ones discussed in Appendix \ref{App:int}. Thus, using \eqref{trueP} and \eqref{cdfSIC_CF} in \eqref{opSIC_CF} with $l=1$ completes the derivation of the \ac{OP} in \ac{SIC} decoding for best user.
%
\section{Asymptotic Analysis of Outage Probability}
\label{Sec:AsympOP}
This section presents the asymptotic analysis to obtain simplified expressions of \ac{OP} for both \ac{CM} and \ac{SIC}-based decoding schemes under the assumption that $\eta_u,\eta_s \rightarrow \infty$.
\subsection{Asymptotic OP for CM-based decoding}
Similar to the approach adopted in Section \ref{OPCapture}, the asymptotic \ac{CDF} $F_{\gamma_{us}}^{\infty}(x)$ and the \ac{MGF} $M_{\gamma_{us}}^{\infty}(-t)$ can be written as
\begin{align}
     F_{\gamma_{us}}^{\infty}(x) &= 1 - \sum\limits_{(k,p,q)}C(k,p,q) \, \left( 1 + \frac{\bar{p}}{\mathcal{I}\,\eta_{us}}\right) x^{\bar{p}}, \label{cdfAsympCM} \\
     M_{\gamma_{us}}^{\infty}(-t) &= 1 - \sum\limits_{(k,p,q)}C(k,p,q) \, \bar{p}\,! \, \left( 1 + \frac{\bar{p}}{\mathcal{I}\,\eta_{us}}\right) \, t^{-\bar{p}},
 \end{align}
where
\begin{align}
     C(k,p,q) & = \frac{(-1)^q \, 2\, k! \, \zeta(k) \, \alpha_{us} \, \mathcal{I}^{\bar{p}}}{p! \, q! \, (\alpha \,\bar{p}+2)\, (\beta - \delta)^{(k+1-\bar{p})}} \nonumber \\
     & \quad \times \left(\frac{r_{\textup{max}}^{(\alpha \,\bar{p}+2)} - r_{\textup{min}}^{(\alpha \,\bar{p}+2)}}{r_{\textup{max}}^2 - r_{\textup{min}}^2}\right), \nonumber \\
     \mathcal{I} &= (U-1)\, \mathbb{E}[r_{is}^{-\alpha}\,|\hat{h}_{us}|^2] + U\, \sigma^{2}_{e_{us}}\, \mathbb{E}[r_{us}^{-\alpha}], \nonumber \\
     a &= \eta_u \, \mathcal{I} + 1 \text{ and }\bar{p} = p + q. \nonumber
 \end{align}
\begin{algorithm}[t!]
\caption{OP simulation for CM and SIC}\label{algo}
    \begin{algorithmic}
    \State \textbf{Initialize:} No. of channel realizations L
    \State \textbf{Initialize:} $U, S, r_{\text{min}}, \theta_0$, and other link parameters
    \State \textbf{Generate:} Channel $\mathbf{H}_{U,S,\text{L}}$, $\mathbf{G}_{S,\text{L}}$, Range $\mathbf{R}_{U,S,\text{L}}$
        \For{$l = 1 \text{ to } U$}
            \State \textbf{Compute}: $\gamma_{us}^{(l)} \quad \forall \, u \notin \mathbf{D}[\,l-1\,]$
            \State \textbf{Perform:} MRC:  $\gamma_u^{(l)} \gets \sum_{s=1}^{S} \gamma_{us}^{(l)} \quad \forall \, u \notin \mathbf{D}[\,l-1\,].$
            \If {$l=1$}
                \State CM: $\gamma_u \gets \texttt{sort}\, \gamma_u^{(1)}$
                \State SIC: $\gamma_{\mathbf{D}\{l\}}^{(l)} \gets \max\limits_{u} \gamma_{u}^{(l)}$
            \ElsIf {$l>1 \, \& \, \gamma_{\mathbf{D}\{l-1\}}^{(l-1)} > \gt$}
                \State SIC: $\gamma_{\mathbf{D}\{l\}}^{(l)} \gets \max\limits_{u, u \notin \mathbf{D}[\,l-1\,]} \gamma_{u}^{(l)}$
            \Else
                \State SIC: $\gamma_{\mathbf{D}\{l\}}^{(l)} \gets 0$
            \EndIf
            \State \textbf{Remove:} Entries of $\mathbf{D}[l]$ from $\mathbf{H}$, $\mathbf{G}$ and $\mathbf{R}$    
        \EndFor
        \State \textbf{Compute}: $\p[K_{\text{vis}} \ge S] = 1-\sum_{j=0}^{S-1} {K \choose j}\, \mathcal{P}^j\, (1-\mathcal{P})^{K-j}$
        \vspace{1pt}
        \State \textbf{Compute}: $\p\left[\gamma_u \le \gt \right] = \texttt{sum}(\gamma_u^{(l)} \le \gamma_{\text{th}})/L$
        \State \textbf{Compute}: $P_{\text{out}}(\mathcal{R}) = \p\left[\gamma_u \le \gt \right] \times \p[K_{\text{vis}} \ge S]$
    \end{algorithmic}
\end{algorithm}
A detailed derivation of the above expressions is presented in Appendix \ref{AssympDerivation}. These expressions are much simpler to comprehend and do not include any integrals. The asymptotic \ac{CDF} $F_{\gamma_{u}}^{\infty}(x)$ and consequently the asymptotic \ac{OP}  $P_{\text{out}}^{\infty}(\mathcal{R})$ in scenarios with multiple satellites can therefore be written as
 \begin{align}
    F_{\gamma_{u}}^{\infty}(x) &= \mathcal{L}^{-1}\left\{\frac{(M_{\gamma_{us}}^{\infty}(-t))^S}{t}\right\}\left(x\right), \nonumber \\
    P_{\text{out}}^{\infty}(\mathcal{R}) &= \left( F_{\gamma_{u}}^{\infty}(\gt) \times \p[K_{\text{vis}} \ge S]\right) + \p[K_{\text{vis}} < S] . \label{OPasympCM}
\end{align}
It can be observed that at high \ac{SNR}, the \ac{OP} attains a floor. It can be attributed to the fact that at high \ac{SNR}, the performance is limited by the interference and the mismatch due to imperfect \ac{CSI} such that any further increase in \ac{SNR} cannot decrease the \ac{OP}.

The simplified expressions can be used to obtain the optimal number of supported users and the required number of satellites with assured visibility (i.e. the number of satellites $S$ such that $\overline{P}_{\text{vis}}(S) = 1$) for a target \ac{OP}. The optimization problem can be formulated as
\begin{align}
    \max \quad & S \nonumber\\
    \text{s.t.} \quad & \overline{P}_{\text{vis}}(S) = 1,\, \theta_0 \in [0,90], \nonumber\\
    & S \in [1,K],\, S \in \mathbb{Z}^+. \nonumber
\end{align}
Using the optimized values of satellites $S^*$ and mask angle $\theta_0^*$, the number of supported users can be obtained by solving
\begin{align}
    \max \quad & U \nonumber\\
    \text{s.t.} \quad & P_{\text{out}}^{\infty}(\mathcal{R}) \leq 0.001 ,\nonumber\\
    & U \geq 1,\, U \in \mathbb{Z}^+. \nonumber
\end{align}
\txtblue{The above optimization problems can be solved using a \ac{GA} inspired by the natural selection process. It finds optimal or near-optimal solutions by iteratively applying genetic operators such as selection, crossover, and mutation to evolve the population over successive generations. If $G$ represents the maximum number of generations, $N$ represents the initial size of the population, and $L$ represents the size of the chromosome (population of candidate solutions), then the upper bound on the time complexity of the entire GA over all iterations can be approximated as $\mathcal{O}(G(N + NL))$.} The GA can solve both constrained and unconstrained problems with linear, non-linear and integer constraints. However, for integer constraints, the \ac{GA} can only find solutions with non-linear inequalities. Hence the assured visibility condition can be approximated and reformulated as $0.999-\overline{P}_{\text{vis}}(S) \leq 0$. As shown in Section \ref{Results}, the above simplified expressions can be used to derive interesting insights on the optimal region of operation in terms of the number of devices, satellites, mask angle and constellation size.

{\renewcommand{\arraystretch}{1}
\begin{table}[t!]
\centering \normalsize
\caption{List of parameters considered for simulation}
\label{params}
\begin{tabular}{lll}
\hline \hline
\textbf{Parameter} & \textbf{Value} & \textbf{Ref.} \\
\hline
Mask elevation angle $\theta_0$ & $10^{\circ}$ &  \multirow{5}{*}{\cite{standard_1}} \\
Target rate $\mathcal{R}$ & $10$ kbps &  \\
Bandwidth $B$ & $125$ kHz &  \\
User antenna transmit gain & 0 dBi & \\
Satellite antenna Tx/Rx gain & 30 dBi & \\
\hline
Constellation Size $K$ & $720$ & \multirow{5}{*}{\cite{niloofarDownlink}} \\
Constellation altitude $h$ & $1200$ km &  \\
Radius of the Earth $r_e$ & $6371$ km &  \\
Noise power at the GS $\sigma_w^2$ & $- 98$ dBm &  \\
Path loss exponent $\alpha$ & $2$ &  \\
\hline
Numerical Laplace Inverse & $D=10$, $Q=15$ & \multirow{2}{*}{\cite{laplaceInv}} \\
& $\text{ln}(10), N=21$ & \\
\hline
Average Shadowing ($m,b,\omega$) & ($2, 0.063, 0.0005$) & \multirow{2}{*}{\cite{ch_value}} \\
Heavy Shadowing ($m,b,\omega$) & ($5, 0.251, 0.279$) &  \\
\hline \hline
\end{tabular}
\end{table}
}
\textbf{Remark (diversity order):} Although the \ac{OP} attains a floor in scenarios with interference and imperfect \ac{CSI}, the diversity order can be determined for a simpler scenario with perfect \ac{CSI} and no interference from other users. The \ac{OP} at high \ac{SNR} can be approximated as
\begin{equation}
P_{\text{out}}^\infty(\mathcal{R}) = (G_c\eta_{us})^{-d} + O(\eta_{us}^{-d}),
\label{op_div}
\end{equation}
where $G_c$ and $d$ denote the cooperation gain and the diversity order, respectively. Here, it is worth mentioning that the \ac{CM} and the \ac{SIC} schemes are synonymous with each other under the no-interference scenario. Using \eqref{sinrAsymp} with the approximation of $F_{H_{us}}(x)$ as shown in \cite{cdfssr_asymp}, and ignoring the higher order terms, the asymptotic \ac{OP} under the no-interference scenario can be written as
\begin{align}
    P_{\text{out}}^\infty(\mathcal{R}) &= \frac{(\alpha_{us} \, R_0 \, \gt)^S}{\Gamma(S+1)} \left(\frac{1}{\eta_{us}}\right)^S,\label{divOrder}
\end{align}
where $R_0 = \frac{2}{\alpha+2}\,(\frac{r_{\textup{max}}^{(\alpha+2)} - r_{\textup{min}}^{(\alpha+2)}}{r_{\textup{max}}^2 - r_{\textup{min}}^2})$. Hence the proposed topology achieves a diversity order of $S$ and a cooperation gain $G_c = \sqrt[S]{\Gamma(S+1)}\, (\alpha_{us} \, R_0 \, \gt)^{-1}$. It is also intuitively verifiable since there are $S$ independent paths between every user and the \ac{GS}. Also, the term $\sqrt[S]{\Gamma(S+1)}$ indicates that as the number of satellites ($S$) increases, the cooperation gain $G_c$ grows sublinearly. This suggests that while adding more satellites improves cooperation, the marginal gain diminishes with each additional satellite.
\begin{figure}[t!]
\centering
\includegraphics[width=0.98\columnwidth]{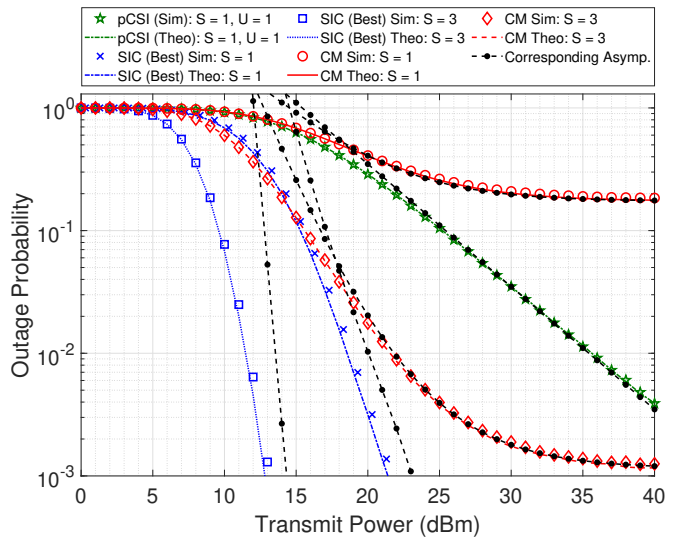}
\caption{\txtblue{Validation of theoretical and simulation results: OP vs transmit power $P_u$ in both SIC and CM for $U=5$ and different number of satellites $S$. Expressions for no-interference (U=1) and perfect CSI (pCSI) are also verified.}}
\label{validation}
\end{figure}
\begin{figure*}[t!]
\centering
\setkeys{Gin}{width=\linewidth}
\begin{tabularx}{\linewidth}{XXXX}
    \centering
     \includegraphics[width=2.4in]{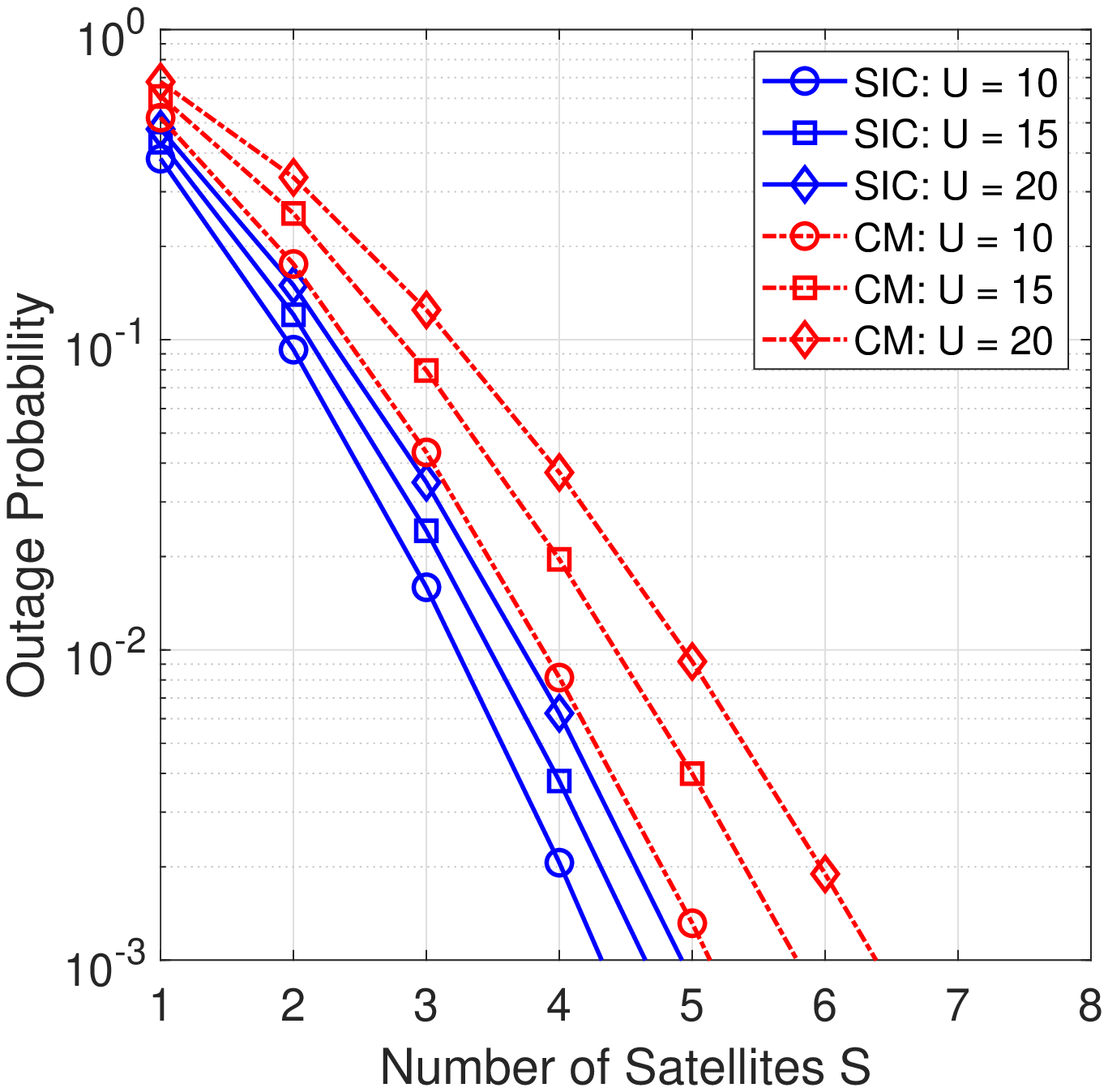}
    \caption{Effect of number of satellites $S$: Averaged \ac{OP} vs $S$ in both SIC and \ac{CM} at $P_u = 20$ dB and different number of users $U$.}
    \label{OPVsSat}
  &
    \centering
    \includegraphics[width=2.45in]{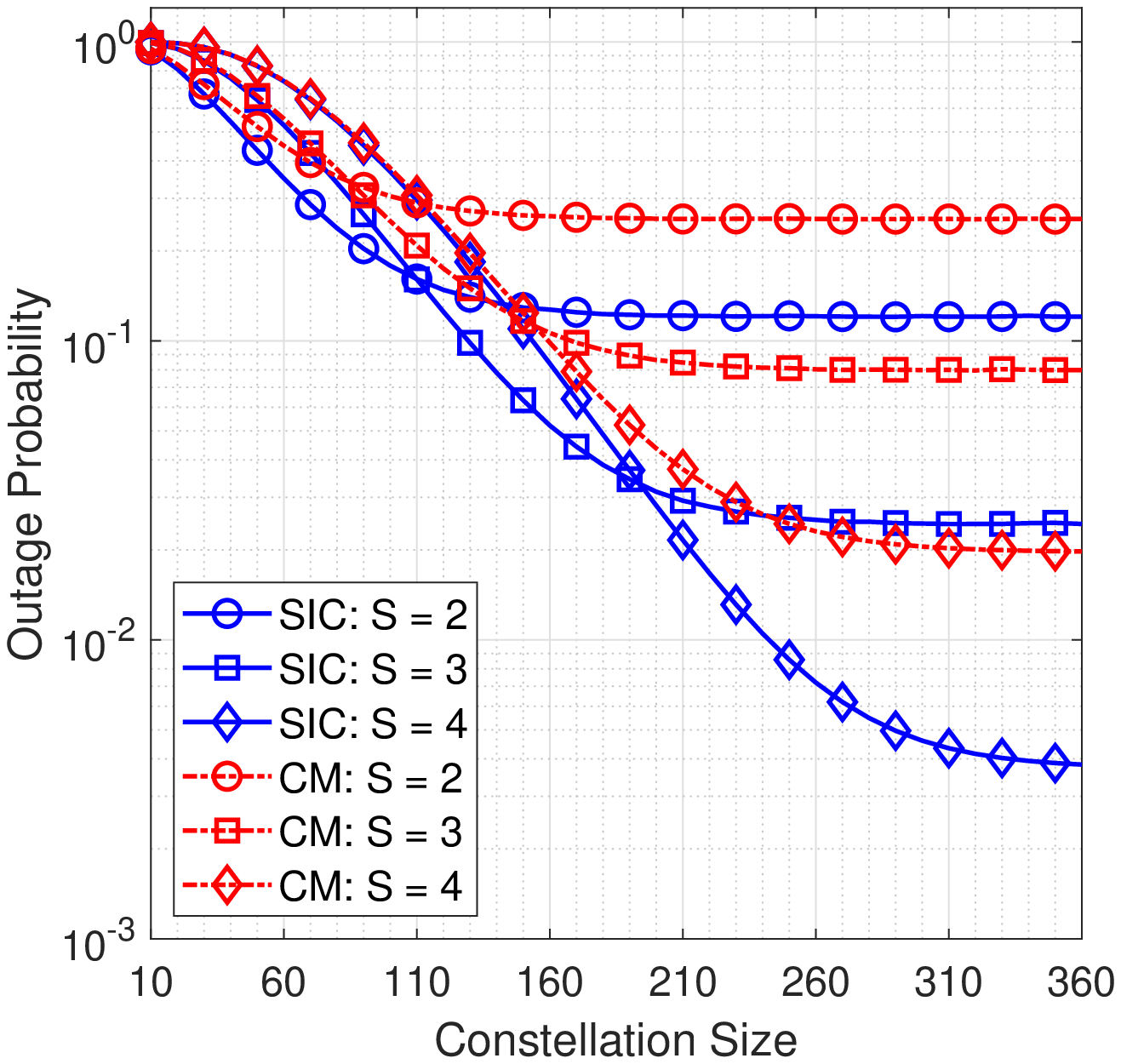}
    \caption{Effect of constellation size $K$: OP vs $K$ for both SIC and CM at $P_u = 20$ dB, $U=15$, and different number of satellites $S$.}
    \label{OPVsConstillationSize}  
\end{tabularx}
\end{figure*}
\subsection{Asymptotic OP for SIC-based decoding}
Since the asymptotic \ac{SINR} $\gamma_{us}^{\infty}$ in \eqref{sinrAsymp} is independent of $G_s$, the asymptotic \ac{SINR} $\gamma_{u}^{\infty(1)}$ and consequently the \ac{OP} $P_{\text{out}}^{\infty (1)}(\mathcal{R})$ for the best user in SIC-based decoding can be conveniently calculated as 
\begin{align}
    \gamma_{u}^{\infty(1)} &= \max\limits_u \left(\sum_{u=1}^{U} \gamma_{us}^{\infty}\right), \nonumber \\
    P_{\text{out}}^{\infty (1)}(\mathcal{R}) &= F_{\gamma_{u}^{(1)}}^{\infty}(\gt) \nonumber\\
    & = \left( \left[F_{\gamma_{u}}^{\infty}(\gt)\right]^U \times \p[K_{\text{vis}} \ge S]\right) + \p[K_{\text{vis}} < S]
    \label{OPasympSIC}.
\end{align}

Similar to \ac{CM}-decoding, the optimal number of supported users and the required number of satellites with assured visibility can be obtained in \ac{SIC} as well.
%
\section{Results}
\label{Results}
This section presents simulation and theoretical results derived in this work to get useful insights into the system. The algorithm for computing OP using Monte-Carlo simulations for both the proposed decoding schemes is provided in Algorithm \ref{algo}. In this section, initially the theoretical analysis is validated with the simulation results. Later the effect of various system parameters on \ac{OP} performance is analyzed. Since many parameters affect the performance, an attempt is made to understand them one by one by keeping all other parameters constant. All the simulations and plots have been generated using MATLAB with $= 10^5$ channel realizations. The default parameters used for simulations, unless stated otherwise are mentioned in Table \ref{params}. The selection of link and stochastic geometry related parameters has been made following 3GPP TR 36.763 \cite{standard_1} and \cite{niloofarDownlink}, respectively. The imperfect \ac{CSI} has been modelled using $\chi = 0.05, \phi = 10$ and $\xi=0.05$ as done in \cite{iCSI2}. Also while computing the numerical Laplace inverse as in \eqref{cdfCM_CF_1}, parameters mentioned in Table \ref{params} are used to maintain the discretization and truncation error less than $10^{-10}$ which is negligible compared to the range of derived \ac{OP}. For solving the optimization problem using \ac{GA} in MATLAB, the \texttt{ConstraintTolerance}, \texttt{FunctionTolerance}, and \texttt{PopulationSize} were set to $10^{-6}$, $10^{-8}$, and $50$ respectively. \txtblue{A desktop PC with Intel(R) Core(TM) i7-8700 CPU operating at 3.20 GHz $\times$ 6 cores and 32 GB of memory was utilized for running the solver. The GA algorithm converged in 70 iterations (median of 100 runs), taking 0.2262 sec per iteration on average to obtain the number of satellites $S^*$ and the mask angle $\theta_0^*$. Given $S^*$ and $\theta_0^*$, it converged in 52 iterations (median), taking 4.4152 sec per iteration on average to obtain the maximum number of users $U^*$}
\subsection{Validation of theoretical and simulation results}\label{sec:val}
Fig. \ref{validation} shows the average \ac{OP} vs transmit power $P_u$ for $U=5$ users in the case of $S=1$ and $S=3$ satellites. The \ac{OP} for every user in the order of their \ac{SINR}s has been calculated and then averaged to obtain the average \ac{OP} of the system. It is observed that the average \ac{OP} derived theoretically using the approximation is very close to the simulation results for \ac{CM} decoding and best user in \ac{SIC} based decoding. This validates the correctness of the derivation presented in Section \ref{OPCapture} and Section \ref{OPSIC}. \txtblue{It can also be observed that the asymptotic curves derived analytically in \eqref{OPasympCM} and \eqref{OPasympSIC} approach the simulated curves rapidly, thus validating the correctness of the derived expressions. Moreover, the asymptotic expression presented in \eqref{divOrder} for the special case with no interference ($U=1$) and perfect CSI (pCSI) also matches with the simulations. It can indeed be observed that the slope of the curves is also equal to the number of satellites used for AF ($S=1$ in this case), thus validating the diversity order as well.} Only simulation results are presented in further results to maintain the brevity of the paper.

Two more observations can be made from Fig. \ref{validation}. First, the \ac{OP} per user decreases with an increase in transmit power of the \ac{IoT} users. An \ac{IoT} user's feasible transmit power range from 12 dB to 20 dB can achieve \ac{OP} ranging from $10^{-1}$ to $10^{-3}$ in \ac{SIC}. Second, as the transmit power increases, the interference effect starts dominating, thus leading to the performance difference between \ac{SIC} and the \ac{CM} decoding. However, with an increase in the number of satellites, the \ac{OP} decreases sharply. \txtblue{The scenario with $S=1$ represents the conventional satellite communication scheme without multiple satellites being visible, unlike the mega-\ac{LEO} constellations. At high \ac{SNR}s, the \ac{OP} in the case of \ac{CM} decoding decreases from $10^{-1}$ to $10^{-3}$ by the addition of two more satellites.} It can be observed that by leveraging the benefits of multiple visible satellites, transmit power of 30-35 dB in \ac{CM} and 12-15 dB for the best user in \ac{SIC} can achieve an \ac{OP} of $10^{-3}$ in a 3 satellite, 5 user system.
\begin{figure*}[t!]
\centering
\setkeys{Gin}{width=\linewidth}
\begin{tabularx}{\linewidth}{XXXX}
    \centering
    \includegraphics[width=2.4in]{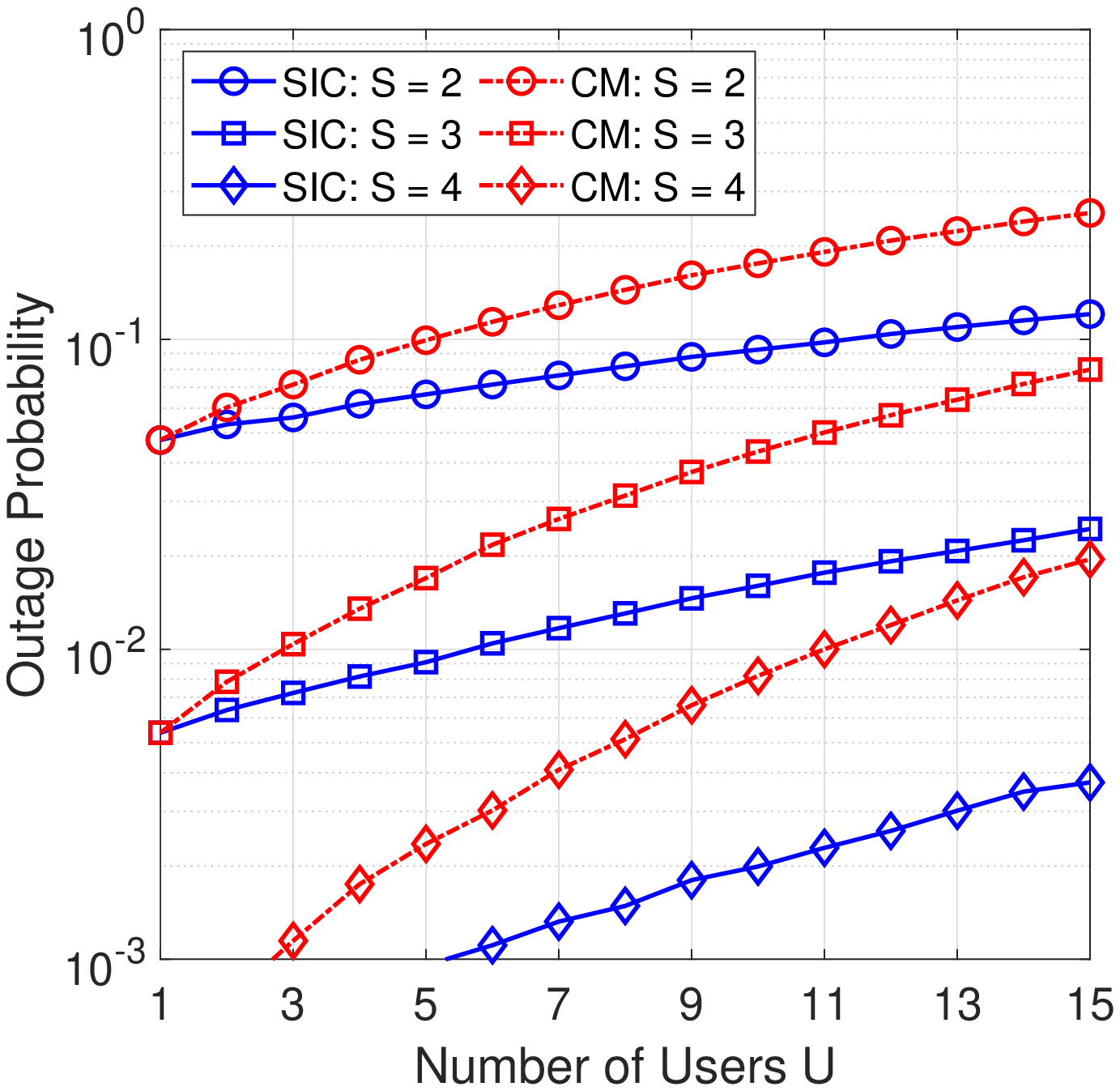}
    \caption{Effect of number of users $U$: Average \ac{OP} vs $U$ for both \ac{SIC} and \ac{CM} scheme at $P_u = 20$ dB and different values of satellites $S$.}
    \label{OPVsUsers}
  &
    \centering
    \includegraphics[width=2.4in]{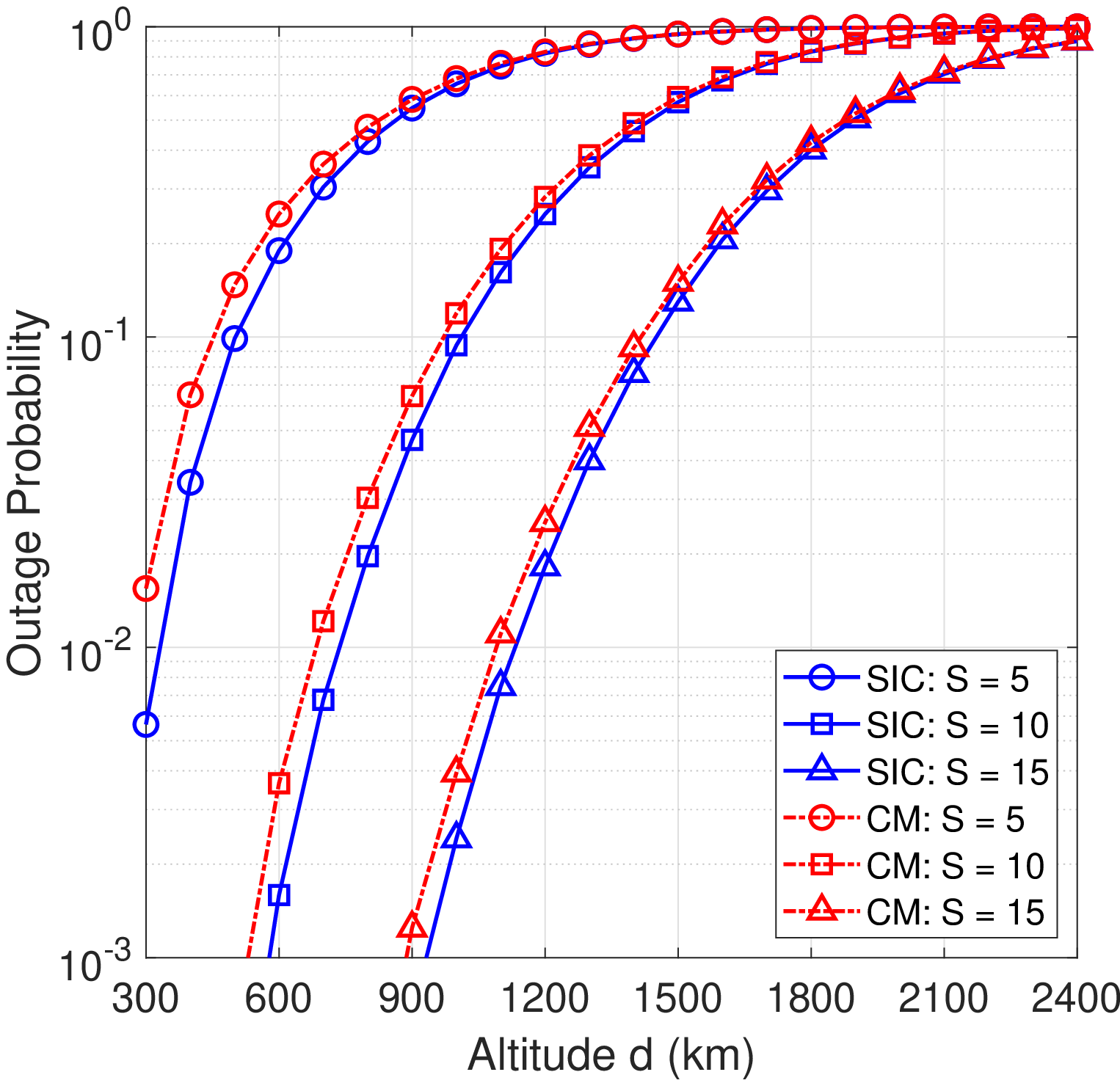}
    \caption{Effect of altitude $d$: Average \ac{OP} vs $d$ for both SIC and \ac{CM} at $P_u = 20$ dB, $U=15$ and different values of satellites $S$.}
    \label{OPVsAltitude}
\end{tabularx}
\end{figure*}
\begin{figure*}[t!]
\centering
\setkeys{Gin}{width=\linewidth}
\begin{tabularx}{\linewidth}{XXXX}
    \centering
    \includegraphics[width=2.4in]{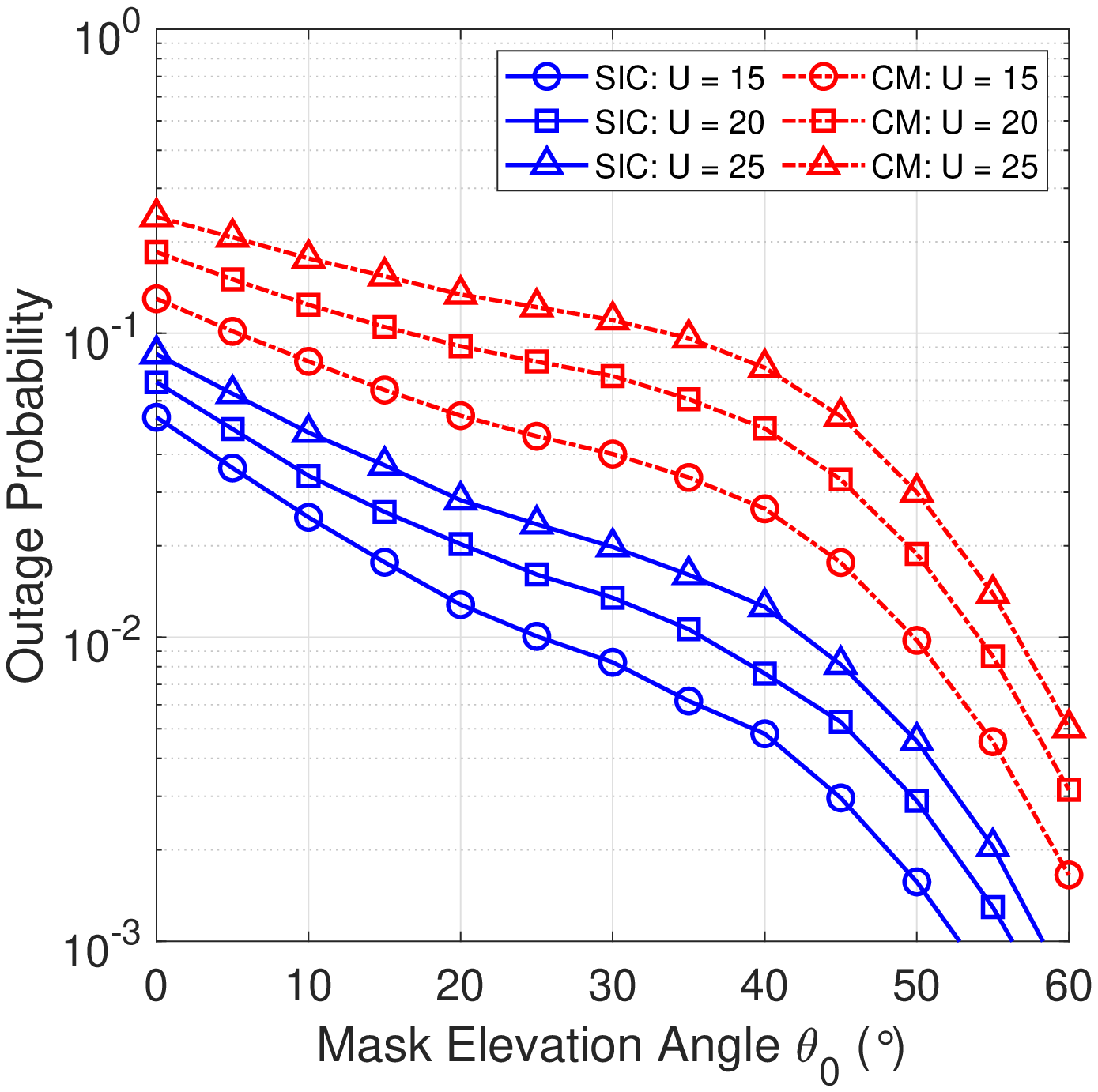}
    \caption{Effect of mask elevation angle $\theta_0$: \ac{OP} vs $\theta_0$ for both \ac{SIC} and \ac{CM} at $P_u = 20$ dB, $S=3$ and different number of users $U$.}
    \label{OPVsEl}
  &
    \centering
    \includegraphics[width=2.4in]{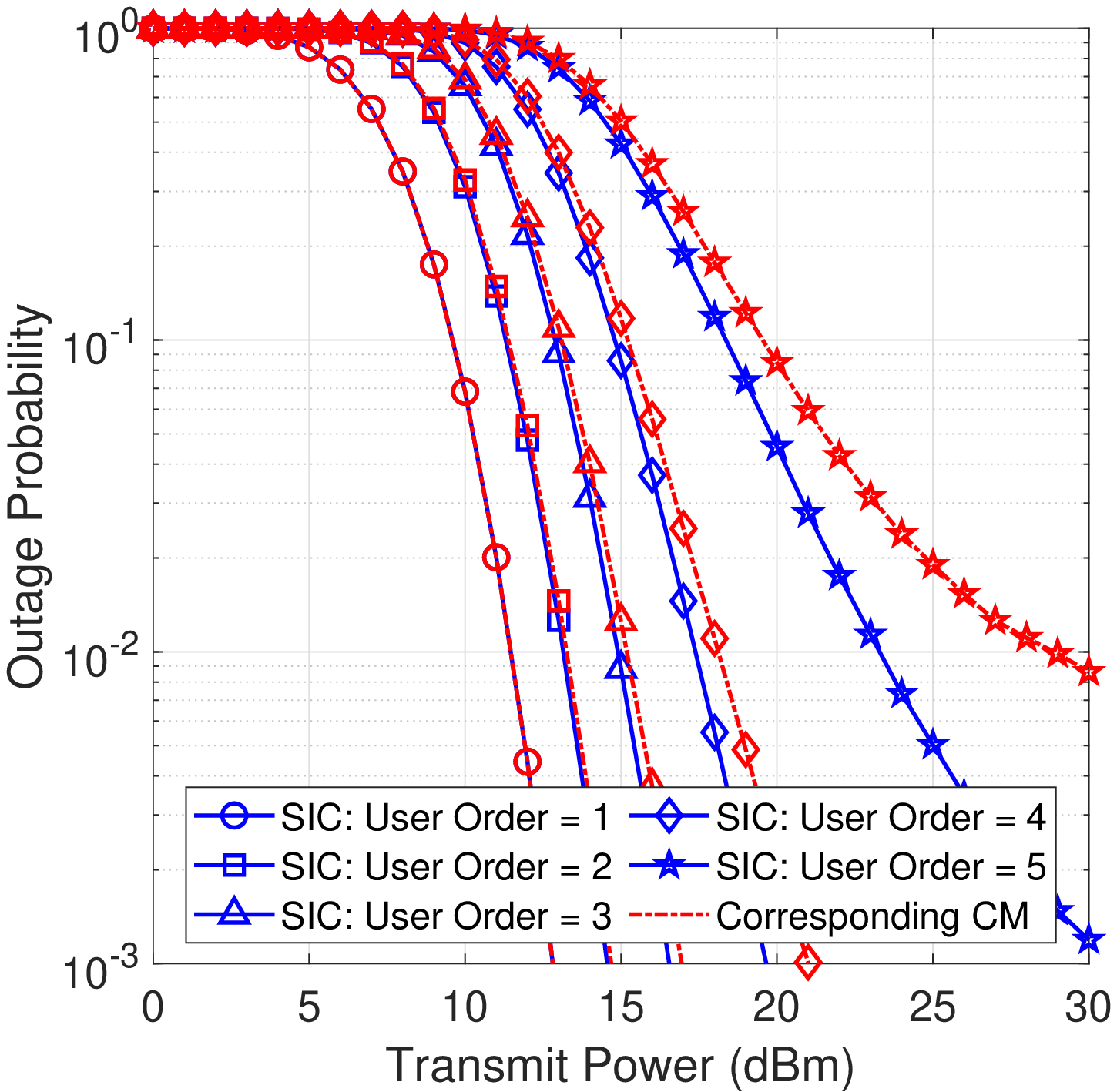}
    \caption{Effect of decoding order $l$: \ac{OP} vs transmit power $P_u$ for ordered decoding in both \ac{SIC} and \ac{CM} for $U=5$ and $S=3$.}
    \label{OPVsPtOrdered}
\end{tabularx}
\vspace{-0.2cm}
\end{figure*}
\subsection{Effect of the number of satellites $S$}
Fig. \ref{OPVsSat} shows the \ac{OP} as a function of number of satellites $S$ for $U=10, 15 \text{ and } 20$ users at $P_u=20$ dBm. It is observed that for a fixed number of users, as the number of satellites increases, the \ac{OP} decreases sharply. It can be observed that merely an addition of 3-4 satellites can reduce the \ac{OP} from $10^{-1}$ to $10^{-3}$ in both \ac{CM} and \ac{SIC} decoding. This also clearly demonstrates how the \ac{IoT} users can leverage multiple visible satellites of the constellation to enhance system performance. Additionally, it is interesting to note that the \ac{OP} for 15 users case in \ac{SIC} is less than the \ac{OP} for 5 users case in \ac{CM}. 
\txtblue{
\subsection{Effect of constellation size $K$}
Fig. \ref{OPVsConstillationSize} shows the effect of the constellation size $K$ on OP for $U=15$ users at $P_u = 20$ dB, and $S=2, 3 \text{ and } 4$ satellites. It can be observed that OP decreases smoothly with an increase in $K$ until a floor is reached. In this case, the floor represents the scenario where OP can't be decreased further, even by adding more satellites to the constellation. The point at which OP saturates denotes the constellation size for which the visibility of $S$ satellites to be utilized in AF-relaying can be ensured (i.e. the constellation size such that $\p[K_{\text{vis}} \ge S] = 1$). Given a target OP to be achieved and the number of satellites intended to be used for AF-relaying, this figure can be utilized to obtain the minimum required size of the mega-LEO constellation. For example, if 4 satellites are to be used for AF-relaying, increasing the constellation size from 150 to 360 can improve the OP from $10^{-1}$ to $3*10^{-3}$ in the case of SIC decoding and from $10^{-1}$ to $2*10^{-2}$ in case of CM decoding. It can also be observed that for a fixed $S$, the performance difference between SIC and CM increases with an increase in $K$ before saturating. This can be attributed to the fact that the term $\p[K_{\text{vis}} < S]$ in \eqref{opCM} and \eqref{opSIC_CF}  dominates at low $K$ values irrespective of the decoding schemes. However, as $K$ increases, $\p[K_{\text{vis}} \ge S]$ also increases, making the performance of decoding schemes more evident.}

\subsection{Effect of the number of users $U$}
Fig. \ref{OPVsUsers} shows the impact on the \ac{OP} as a function of number of users $U$ at $P_u=20$ dBm for $S=2, 3 \text{ and } 4$. It can be observed that the \ac{OP} increases with an increase in the number of users due to the increase in interference. It can also be observed that the performance gap between the \ac{SIC} and \ac{CM} also increases with an increase in $U$. This can be attributed to the fact that the impact of interference decreases with decoding of subsequent users in \ac{SIC} whereas \ac{CM} assumes a constant number of interferers for all the users. It is interesting to note that the difference between the performance of \ac{SIC} and \ac{CM} becomes significant as more and more satellites are added to the system. \looseness=-1
\begin{figure*}[t!]
\centering
\setkeys{Gin}{width=\linewidth}
\begin{tabularx}{\linewidth}{XXXX}
    \centering
    \includegraphics[width=2.43in]{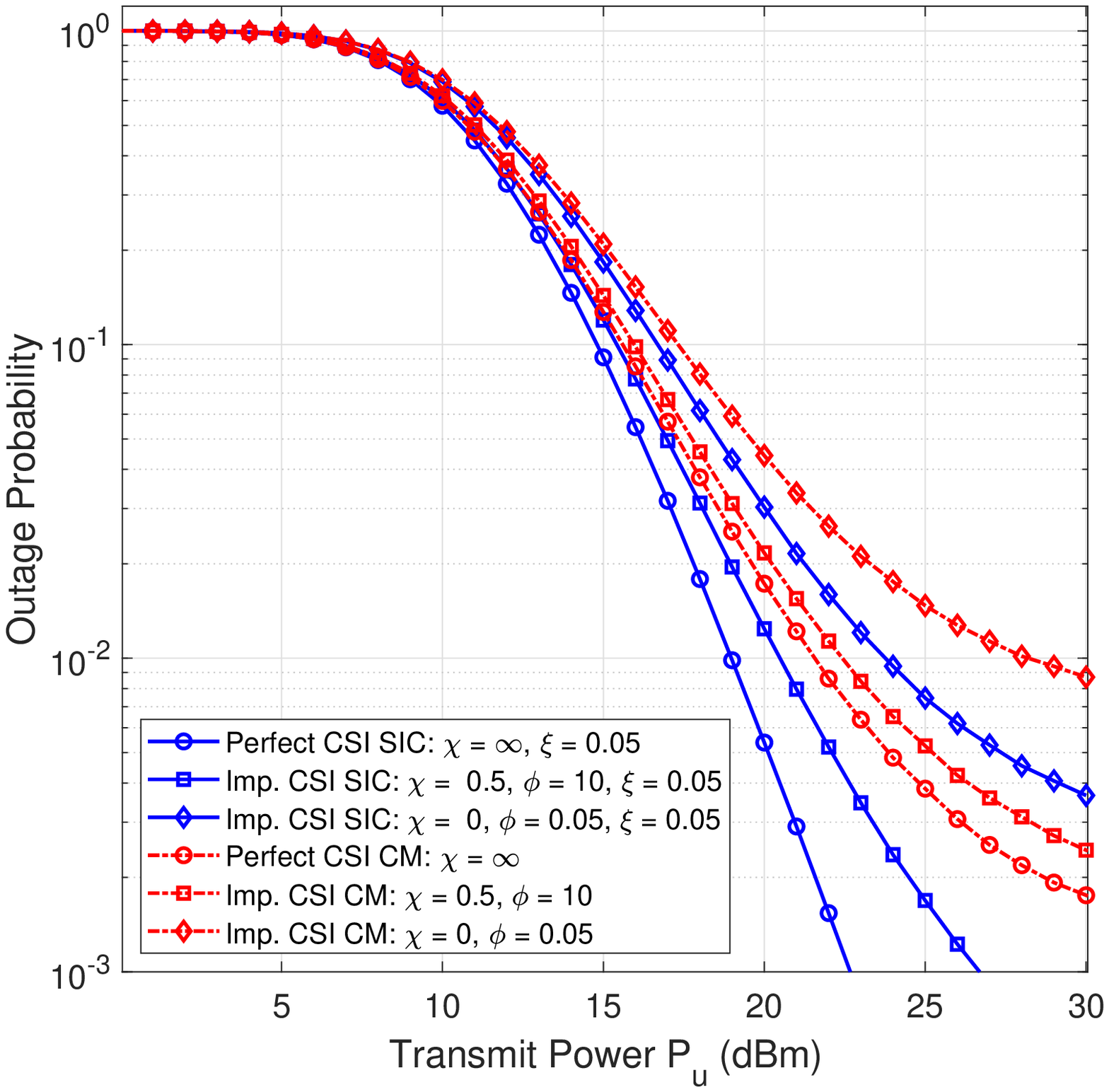}
    \caption{Effect of imperfect CSI: OP vs SNR plots for scenarios with perfect ($\chi=\infty$) and imperfect CSI in both CM and SIC-based decoding for $U=5$, $S=3$ and different values of mismatch parameters $\chi, \phi$ and $\xi$. Here $\chi=0$ represents SNR-independent, and $\chi \neq 0$ represents SNR-dependent CSI mismatch.}
    \label{iCSI}
  &
    \centering
    \includegraphics[width=3.1in,trim={7.9cm 0.2cm 9.8cm 1.4cm},clip]{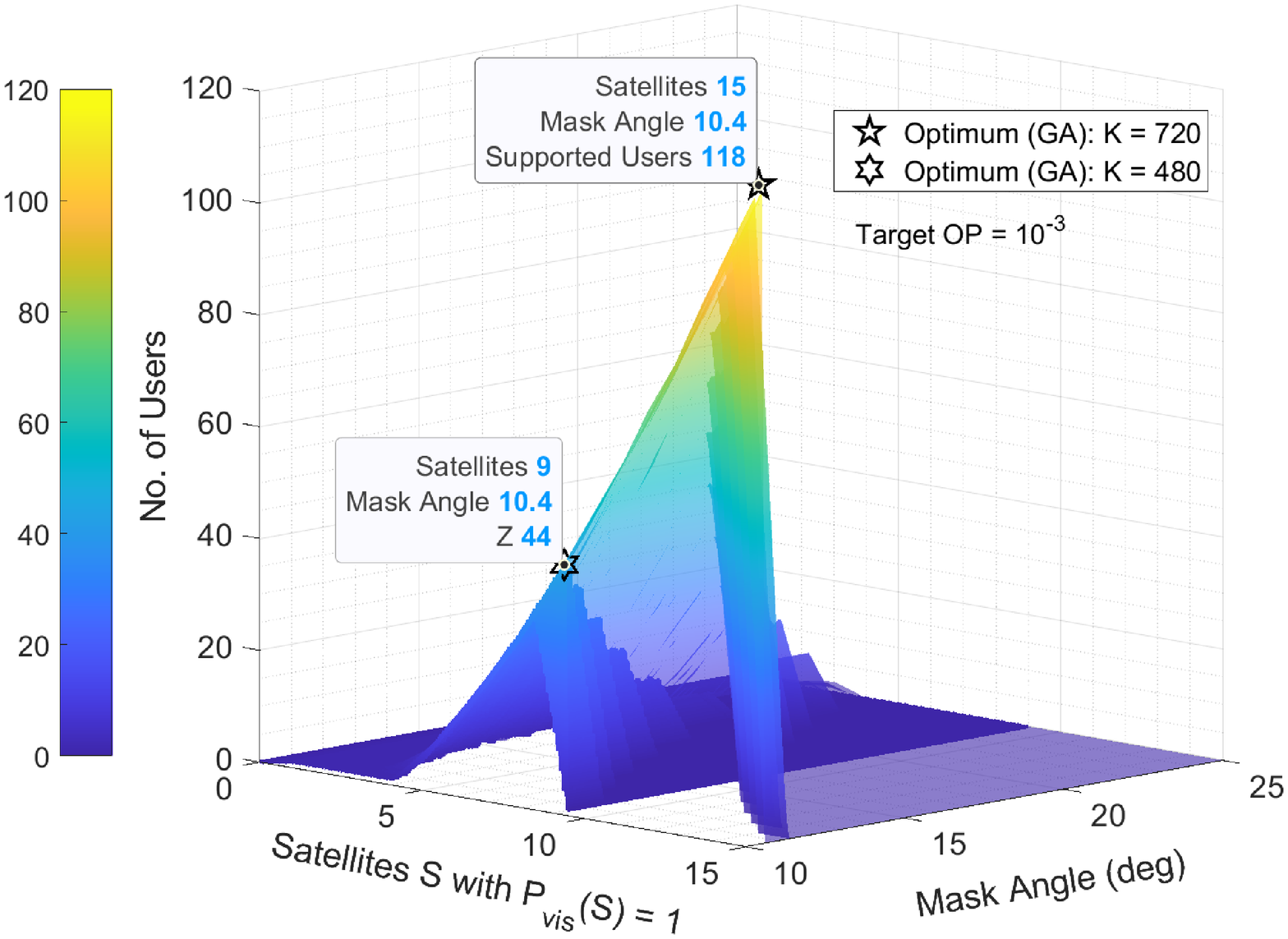}
    \caption{Maximum number of users which can achieve the target \ac{OP} of $10^{-3}$ using \ac{CM} decoding versus the number of satellites S required with $\overline{P}_{\text{vis}}(S) = 1$ (also mentioned as number of satellites with assured visibility) and varying mask angle. (generated for a constellation of 720 satellites at 1200 km and \ac{IoT} users with $P_u$ = 20 dB)}
    \label{UvsSvsM}
\end{tabularx}
\end{figure*}
%
\subsection{Effect of altitude $d$}
Fig. \ref{OPVsAltitude} shows the impact of altitude on the \ac{OP} for $S=5, 10 \text{ and } 15$ satellites and $U=15$ users at $P_u=20$ dBm. It can be observed that \ac{OP} increases as the altitude increases. This is because the constellation's altitude directly impacts the path loss observed by the signals. It can also be observed that while more satellites enhance the \ac{OP} performance because of better elevation angles, the benefit of adding more satellites diminishes rapidly with increasing altitude. Hence, the number of satellites and the selection of decoding scheme can be traded-off with the altitude and number of users.

Usually, the \ac{LEO} satellites are considered to be placed between 600 km to 1800 km. Hence for a desired \ac{OP} at a fixed transmit power, Fig. \ref{OPVsAltitude} can be used to determine the minimum number of satellites required in look angle for various constellations at different altitudes. For example, using \ac{SIC}, in a network of 15 active users, a minimum of 15 satellites at nearly 900 km are required to achieve an \ac{OP} of $10^{-3}$. A similar performance can be achieved with 10 satellites only if placed at an altitude of 600 km. 
%
%
\subsection{Effect of mask elevation angle $\theta_0$}
Fig. \ref{OPVsEl} shows the effect of $\theta_0$ on \ac{OP} for $S=3$ satellites and $U=15, 20 \text{ and } 25$ users at $P_u=20$ dBm. The \ac{OP} decreases with an increase in $\theta_0$; however, the rate of decrease changes at around $\theta_e=40^\circ$ for the shown case of $S=3$. It can be explained by the two-fold impact of $\theta_0$ on the \ac{OP}. As evident from \eqref{rMax} and \eqref{eqn:SVis_2}, $A_{\text{vis}}$ decreases with an increase in $\theta_0$. Hence the probability of seeing a defined number of satellites decreases with an increase in $\theta_0$ for every user. It, therefore, tends to increase the \ac{OP}. On the other hand, as evident from \eqref{avgpL}, an increase in $\theta_0$ decreases the distance and, consequently, the average path loss between the users and the satellites. It, therefore, tends to reduce the \ac{OP}. The impact of reducing path loss dominates nearly till $\theta_0=40^\circ$. However, after that, the reduced probability of seeing a defined number of satellites rapidly decreases the \ac{OP}.
%
\subsection{Effect of decoding order $l$ and trade-offs}
Fig. \ref{OPVsPtOrdered} shows the \ac{OP} for various users in ordered decoding, where user 1 means the first user being decoded in \ac{SIC}. Similarly, in a 5-user scenario, user 5 means the last user decoded in \ac{SIC}. The \ac{OP} for corresponding \ac{CM} has also been shown in Fig. \ref{OPVsPtOrdered} where the user order is solely determined by the \ac{SINR}s without removing interference for any user. It can be observed that the initial few users in the order have similar \ac{OP} for both \ac{SIC} and \ac{CM}. It can be attributed to the fact that the interference in both \ac{SIC} and \ac{CM} remains nearly similar for the initial few users. However, for higher order decoding in \ac{SIC}, the interference decreases significantly due to the subtraction of information signals of the decoded user. This is not the case in \ac{CM}; decoding of higher order users happens in heavy interference from users with better channel conditions. Hence the difference between the \ac{OP} performance of \ac{SIC} and \ac{CM} increases significantly for higher-order users. It can be concluded that trading off with the throughput and desired \ac{OP}, the initial few users can be decoded using \ac{CM} only, thus reducing the decoding complexity. However, for applications requiring decoding of all or most users, \ac{SIC} is preferable to \ac{CM}.
\subsection{Effect of channel imperfections}
Fig. \ref{iCSI} shows the average OP versus the transmit power $P_u$ for both the CM and the SIC-based schemes for $U=5$ users and $S=3$ satellites under both perfect and imperfect CSI scenarios. The results demonstrate that the OP performance deteriorates in the presence of CSI mismatch compared to the ideal case when perfect CSI ($\chi = \infty$) is available. Two types of CSI mismatch: the SNR-dependent CSI mismatch ($\chi \neq 0$), and the SNR-independent CSI mismatch ($\chi=0$) are shown in Fig. \ref{iCSI}. When $\chi \neq 0$, the variance of the CSI estimation error depends on the link SNR, and the impact of $\chi$ is dominant in determining the system performance. The rate of decline in $\sigma^2_{e_{us}}$ and $\sigma^2_{e_{s}}$ depends on $\chi$, and thus an increase in $\chi$ is expected to result in improved OP performance. The SIC decoding is more affected than the CM decoding as the residual error due to imperfect decoding accumulates successively in SIC. For example, at $22$dB transmit power, the OP with imperfect CSI at $\chi = 0.5$ is 3.4 times higher than the OP with perfect CSI in the case of SIC decoding but only 1.3 times higher in the case of CM decoding.

In contrast, when $\chi=0$, the variance of the CSI estimation error solely depends on $\phi$. It can be observed that the impact of imperfect CSI is negligible at low SNRs as the products $\eta_{us},\sigma^2_{e_{us}}$ and $\eta_{us},\sigma^2_{e_{us}}$ approach zero, and the CSI quality tends to be perfect. However, as the transmit power increases, particularly above 15 dBm in SIC and 18 dBm in CM decoding, significant degradation in OP performance is observed.
\subsection{Optimum number of users for a target \ac{OP}}
Fig. \ref{UvsSvsM} shows the approximate number of users which can be served for a target OP of $10^{-3}$ using CM decoding against the number of satellites with assured visibility and varying mask angles for two different constellation sizes, $K=480$ and $K=720$. These results are obtained using the asymptotic expressions derived in Section \ref{Sec:AsympOP}. Here, the number of satellites $S$ with assured visibility refers to the number of satellites required such that $\overline{P}_{\text{vis}}(S) = 1$. It can be observed that $A_{\text{vis}}$ decreases with an increase in $\theta_0$, as derived in \eqref{rMax} and \eqref{eqn:SVis_2}, thereby decreasing the number of satellites with assured visibility. The maximum number of supported users obtained using the \ac{GA} optimizer can be verified using a simple grid-based algorithm where the \ac{OP} is calculated for every possible $U$ for a $(\theta_0,S)$ pair obtained from a grid of all likely $S$, $\theta_0$ values. It is worth mentioning that the asymptotic expressions can be efficiently used for such iterative grid-search methods since they do not contain any integrals to be computed numerically. As shown in Fig. \ref{UvsSvsM}, the optimal values obtained from the \ac{GA} optimizer match the values obtained using the grid search algorithm. For example, considering 720 satellites in the constellation and users with 20 dB transmit power, approximately 118 users can be supported optimally using CM decoding for a target OP of $10^{-3}$ using 15 satellites at $\theta_0 = 10.5^{\circ}$. The number of supported users falls down to 44 if the constellation has 480 satellites only. Similar approximations can also be made for SIC decoding.
%
\section{Conclusion}\label{sec:conc}
This paper presented a simple and energy-efficient direct-access topology for \ac{LEO} satellites based \ac{IoT} networks. The \ac{OP} performance of the topology was analysed using stochastic geometry to model the random satellite locations. In this context, analytical expressions were derived for average \ac{OP} in the \ac{CM} decoding scheme, and the \ac{OP} of the best user was derived for the \ac{SIC} decoding scheme, assuming imperfect knowledge of the \ac{CSI}. Simplified expressions of the \ac{OP} were also derived under high \ac{SNR} assumptions which demonstrate that \ac{OP} attains a floor due to interference and imperfect CSI. The asymptotic expressions were utilized to obtain an optimal region of operation achieving a target \ac{OP}. Although all the \ac{IoT} users transmit at the same power, it was found that the \ac{SIC} decoding scheme performs better than the \ac{CM} decoding scheme. The theoretical analysis was verified through rigorous simulations. The effects of system parameters like the number of users, number of satellites, altitude, and mask elevation angle of the constellation on the \ac{OP} performance were discussed in detail. The results of this paper demonstrate that for the practical values of the above system parameters, the proposed topology is feasible and attractive for low-powered \ac{IoT} networks.

\appendices
\section{Proof for CDF and PDF of \texorpdfstring{$R$}{R}}
\label{proof:range}
The distribution of $R$ can be found in three steps:
\begin{enumerate}
    \item Finding $r_{\text{min}}$, $r_{\text{max}}$ and the relationship between $R$ and the surface area of the spherical cap, $A_{\text{cap}}$ formed by the satellites at the distance less than or equal to $R$
    \item Finding the distribution of $A_{\text{cap}}$
    \item Finding the distribution of $R$ from the distribution of $A_{\text{cap}}$
\end{enumerate}

\textit{Step 1:} From basic geometry as as shown in Fig. \ref{sm}, $r_{\text{min}} = d$ is the orbital altitude, observed at $\theta_e = 90^\circ$. Also, $r_{\text{max}}(\theta_0)$ will be observed at the mask elevation angle $\theta_0$. Hereafter it is written as simply $r_{\text{max}}$ to maintain brevity. Applying law-of-cosines for triangles at $\angle XYZ$ gives,
\begin{equation}\label{eqn:cosineRMax}
    (r_e + r_{\text{min}})^2 = r_e^2 + r_{\text{max}}^2 - 2\, r_e\, r_{\text{max}} \cos\,({90+\theta_0}).
\end{equation}
Solving the quadratic equation \eqref{eqn:cosineRMax} for $r_{\text{max}}$ and considering all the distances to be positive, we get
\begin{equation}
    \label{rMax}
    r_{\text{max}} = \sqrt{(r_e \sin \theta_0)^2 + (r_e + r_{\text{min}})^2 - r_e^2} - r_e\sin\theta_0.
\end{equation}

Further derivation to find the relationship between $A_{\text{cap}}$ and $R$ can be done on similar lines of  \cite{niloofarDownlink}. From Fig. \ref{sm} where $m$ and $n$ are shown, we can write
\begin{align}
    A_{\text{cap}} &= \pi\, (m^2+n^2), \label{eqn:SVis}\\
    R^2 &= (r_{\text{min}}-m)^2 + n^2 .\label{eqn:RVis}
\end{align}
Using \eqref{eqn:SVis} and \eqref{eqn:RVis}, we can obtain
\begin{equation}\label{eqn:RVis_2}
    R^2 = r_{\text{min}}^2 - 2\, r_{\text{min}}\,(r_e+r_{\text{min}})(1-\cos\psi) + \frac{A_{\text{cap}}}{\pi}.
\end{equation}
Using the geometric identity for surface area of any spherical cap, we can write $A_{\text{cap}} = 2\pi\,(r_e+r_{\text{min}})^2\,(1-\cos \psi)$. Substituting $A_{\text{cap}}$ in \eqref{eqn:RVis_2}, we get
\begin{align}
    R^2 &= r_{\text{min}}^2 + \frac{A_{\text{cap}}}{\pi} \left( 1 - \frac{r_{\text{min}}}{r_e+r_{\text{min}}} \right).
    \label{eqn:RVis_3}
\end{align}

Similarly at $\theta_e = \theta_0$, a spherical cap of surface area $A_{\text{vis}}$ is formed by the all the visible satellites at distance less than or equal to maximum distance $r_{\text{max}}$ from the user. Hence similar to \eqref{eqn:RVis_3}, for $R=r_{\text{max}}$, we can write
\begin{equation}
    r_{\text{max}}^2 = r_{\text{min}}^2 + \frac{A_{\text{vis}}}{\pi} \left( 1 - \frac{r_{\text{min}}}{r_e+r_{\text{min}}} \right),
    \label{rAvis}
\end{equation}
or
\begin{align}
    A_{\text{vis}} &= \frac{\pi}{r_e} \,(r_{\text{max}}^2-r_{\text{min}}^2)\,(r_e+r_{\text{min}}).\label{eqn:SVis_2}
\end{align}

\textit{Step 2:} From Fig. \ref{sm}, the \ac{CDF} of the surface area of the spherical cap formed by satellites at a random distance less than or equal to $R$ from a user is
\begin{align}
F_{A_{\text{cap}}}(x) = \frac{x}{A_{\text{vis}}}.
\end{align}

\textit{Step 3:} Using the relationship between $A_{\text{cap}}$ and $R$ as derived in \eqref{eqn:RVis_3}, the \ac{CDF} of the distance $F_R(r)$ can be written as
\begin{align}
    F_R(r) &= \mathbb{P}(R \le r) = \mathbb{P}(R^2 \le r^2) \nonumber\\
    & = \mathbb{P}\left(r_{\text{min}}^2 + \frac{A_{\text{cap}}}{\pi} \left( 1 - \frac{r_{\text{min}}}{r_e+r_{\text{min}}} \right) \le r^2\right) \nonumber \\
    &= \mathbb{P}\left(A_{\text{cap}} \le \frac{\pi\,(r^2-r_{\text{min}}^2)}{1 - \frac{r_{\text{min}}}{r_e+r_{\text{min}}}} \right) \nonumber \\
    &= \frac{\pi\,(r^2-r_{\text{min}}^2)}{A_{\text{vis}}\,(1 - \frac{r_{\text{min}}}{r_e+r_{\text{min}}})}. \label{eqn:cdfR_1}
\end{align}
Using \eqref{eqn:SVis_2} in \eqref{eqn:cdfR_1}, we can write
\begin{equation}\label{eqn:cdfR_2}
    F_R(r) = \frac{r^2 - r_{\text{min}}^2}{r_{\text{max}}^2 - r_{\text{min}}^2}, \text{ for } r_{\text{min}} \le r \le r_{\textup{max}},
\end{equation}
where the $r_{\text{max}}$ is given by \eqref{rMax}.
The corresponding \ac{PDF} can be found by differentiating \eqref{eqn:cdfR_2} with respect to $r$. 
\section{Proof for success probability of $K_{\text{vis}}$}
\label{proof:numVisible}
The success probability is given by the ratio of the surface area of the spherical cap region where a visible satellite can lie to the total surface area of the sphere. It can be written as
\begin{equation}\label{PVis_1}
    \mathcal{P} = \frac{A_{\text{vis}}}{4 \pi\, (r_e + r_{\text{min}})^2}.
\end{equation}
Using \eqref{eqn:SVis_2} in \eqref{PVis_1}, we can write
\begin{align}
    \mathcal{P} &= \frac{\pi (r_{\text{max}}^2-r_{\text{min}}^2)(r_e + r_{\text{min}})}{4\pi \,r_e\, (r_e + r_{\text{min}})^2} \nonumber\\
    &= \frac{r_{\textup{max}}^2 - r_{\textup{min}}^2}{4\, r_e\, (r_e + r_{\textup{min}})}.
\end{align}
\section{Derivation of \texorpdfstring{$\mathbb{E}[r_{us}^{-\alpha}\, H_{us}]$}{E{rH}}}
\label{proof:ExpectationRH}
Since $H_{us}$ and $R_{us}$ are independent, $\mathbb{E}\,[r_{us}^{-\alpha}\, H_{us}]$ can be written as
\begin{align}
    \mathbb{E}[r_{us}^{-\alpha}\, H_{us}] &= \mathbb{E}[r_{us}^{-\alpha}]\, \mathbb{E}[H_{us}] .
\end{align}
Using \eqref{pdf_sr} and \eqref{pdfR}, $\mathbb{E}[r_{us}^{-\alpha}]$ and $\mathbb{E}[H_{us}]$ can be solved as
\begin{align}
    \mathbb{E}[r_{us}^{-\alpha}] &= \int\limits_{r_{\text{min}}}^{r_{\text{max}}} r_{us}^{-\alpha}\, f_{R}(r_{us}) \, dr_{us} \nonumber\\
    &=
    \begin{cases}
        \dfrac{2\,(r_{\textup{max}}^{2-\alpha} - r_{\textup{min}}^{2-\alpha})}{(2-\alpha)(r_{\textup{max}}^2 - r_{\textup{min}}^2)} & \text{ for } \alpha \neq 2,\\[12pt]
        \dfrac{2}{r_{\textup{max}}^2 - r_{\textup{min}}^2} \ln\left(\dfrac{r_{\text{max}}}{r_{\text{min}}}\right) & \text{ for } \alpha=2,
    \end{cases}\label{avgpL}
\end{align}
and
\begin{align}
    \mathbb{E}[H_{us}] &= \int\limits_0^\infty h_{us} \, f_{H_{i}}(h_{us}) \,dh_{us} \nonumber\\
    &=\sum\limits_{\kappa = 0}^{m_i-1} \frac{\alpha_i\,\zeta(\kappa)\,\eta_i\,\Gamma(\kappa+2)}{(\beta_i - \delta_i)^{\kappa + 2}} .      
\end{align}
\section{Numerical Integration in \eqref{mgf_3}}
\label{App:int}
The integral term in \eqref{mgf_3} can be efficiently calculated using the \texttt{vpaintegral} function of MATLAB. It uses the global adaptive quadrature technique and variable precision arithmetic to perform the integration. The speed of the execution can be traded-off with the tolerance value. Consider
\begin{align}
    \texttt{func} &= M_{\gamma_{us} \,|\, r_{us}}\left(-t\right) f_R(r_{us}), \nonumber
\end{align}
where $f_R(r_{us})$ and $M_{\gamma_{us} \,|\, r_{us}}\left(-t\right)$ are given in \eqref{pdfR} and \eqref{mgf_2}, respectively. Then, the integral term in \eqref{mgf_3} can be evaluated using
\begin{align}
\texttt{vpaintegral}(&\texttt{func},\texttt{r},\texttt{rMin},\texttt{rMax},...\nonumber\\
&\texttt{\char13 RelTol\char13,1e-4, \char13 AbsTol\char13,0);}\nonumber
\end{align}
where $\texttt{rMin} = r_{\text{min}}, \texttt{rMax} = r_{\text{max}}$ and the integration is done for a relative tolerance of $10^{-4}$ and the option to set the absolute tolerance is turned off. 
\section{Derivation of Asymptotic CDF and MGF of $\gamma_{us}^{\infty}$}
\label{AssympDerivation}
The end-to-end SINR expression for the CM-based decoding as shown in \eqref{sinr_general} can be approximated under the assumption $\eta_u,\eta_s \rightarrow \infty$ as
\begin{align}
    \gamma_{us}^{\infty} &\approx \frac{ r_{us}^{-\alpha}\, H_{us}}{ \sum\limits_{\substack{i=1 \\ i\neq u}}^U r_{is}^{-\alpha} H_{is} + \sum\limits_{u=1}^U \eta_u r_{us}^{-\alpha} \sigma^2_{e_{us}} + 1},
    \label{sinr_CaptureApprox1}
\end{align}
since at high SNR, the system tends to become limited by interference only. For mathematical tractability, we write $\mathcal{I} = (U-1)\, \mathbb{E}[r_{is}^{-\alpha}\,|\hat{h}_{us}|^2] + U\, \sigma^{2}_{e_{us}}\, \mathbb{E}[r_{us}^{-\alpha}]$ and $a = \eta_u \, \mathcal{I} + 1$. Hence $\gamma_{us}$ can be written as
\begin{align}
    \gamma_{us}^{\infty} &\approx \frac{1}{a} \,r_{us}^{-\alpha}\, H_{us}.
    \label{sinrAsymp}
\end{align}

Therefore, using the theorem of transformation of random variables, the asymptotic CDF of $\gamma_{us}^{\infty}$ can be computed as
\begin{align}
   F_{\gamma_{us}}^{\infty}(x) &= \p\left[\frac{1}{a} \,r_{us}^{-\alpha}\, H_{us} \le x \right] \nonumber \\
   &= \int\limits_{r_{\text{min}}}^{r_{\text{max}}}F_{H_{us}}\left(\frac{a\,x}{r_{us}^{-\alpha}}\right)\, f_{R}(r_{us})\,\, dr_{us} \nonumber \\
   &= F_{\widetilde{H}_{us}}\left(a\,x\right), 
   \label{cdf_cm_asymp}
\end{align}
where $F_{\widetilde{H}_{us}}(z)$ is derived in closed form in \eqref{cdfHBar_2}.

By using the series expansion of the lower incomplete Gamma function, $\gamma(\cdot,\cdot)$ as given in \cite[Eq. 8.354.1]{formula} in the above equation, we get
\begin{align}
    & F_{\widetilde{H}_{us}}(z) \nonumber \\
    & = 1 - \left\{\sum\limits_{(k,p,q)} \, \frac{(-1)^q \, 2\, k! \, \zeta(k) \, \alpha_{us}}{p! \, q! \, (\alpha \,\bar{p}+2)\, (\beta - \delta)^{(k+1-\bar{p})}} \, \right. \nonumber \\
    & \left. \qquad \quad \times \left(\frac{r_{\textup{max}}^{(\alpha \,\bar{p}+2)} - r_{\textup{min}}^{(\alpha \,\bar{p}+2)}}{r_{\textup{max}}^2 - r_{\textup{min}}^2}\right) \, (z\,\mathcal{I})^{\bar{p}} \, \left( 1 + \frac{\bar{p}}{\mathcal{I}\,\eta_{us}}\right)\right\} \nonumber \\
    & \quad + \mathcal{O}\left(\frac{1}{\eta_{us}}\right),
    \label{cdfHBarAsymp}
\end{align}
where $q$ is the iterator for the series expansion of $\gamma(\cdot,\cdot)$, $\bar{p} = p + q$ and $\mathcal{O}\left(\frac{1}{\eta_{us}}\right)$ represents the higher order terms of $\left(1/\eta_{us}\right)$. Since $\eta_u,\eta_s \rightarrow \infty$, the \eqref{cdfHBarAsymp} can further be simplified by neglecting the higher order terms to finally obtain \eqref{cdfAsympCM}.

\bibliographystyle{IEEEtran}
\bibliography{IEEEabrv,ref.bib}{}
\end{document}